\documentclass[10pt,conference]{IEEEtran}
\usepackage{cite}
\usepackage{amsmath,amssymb,amsfonts}
\usepackage{algorithmic}
\usepackage{graphicx}
\usepackage{textcomp}
\usepackage{xcolor}
\usepackage[hyphens]{url}
\usepackage{fancyhdr}
\usepackage{hyperref}
\usepackage[export]{adjustbox}
\usepackage{multirow}
\usepackage{booktabs}
\usepackage{pifont}
\usepackage{listings}
\usepackage{threeparttable}
\usepackage{hyperref} 
\usepackage{color}
\usepackage{mathrsfs}
\usepackage{amsmath,amssymb,amsfonts}
\definecolor{codecyan}{rgb}{0,0.6,0}
\definecolor{codegray}{rgb}{0.5,0.5,0.5}
\definecolor{codepurple}{rgb}{0.58,0,0.82}
\definecolor{backcolour}{rgb}{0.95,0.95,0.92}
\usepackage{threeparttable} 
\usepackage{caption}
\newcommand{\hpcayear}{2025}

\newcommand{\hpcasubmissionnumber}{369}
\title{Exploring the Performance Improvement of Tensor Processing Engines through Transformation in the Bit-weight Dimension of MACs }

\def\hpcacameraready{} % Uncomment to build camera-ready version

\newcommand\hpcaauthors{Qizhe Wu$^1$, Huawen Liang$^1$, Yuchen Gui$^1$, Zhichen Zeng$^{1,2}$, Zerong He$^1$, Linfeng Tao$^1$,\\ Xiaotian Wang$^{1,3}$, Letian Zhao$^1$ Zhaoxi Zeng$^1$, Wei Yuan$^1$, Wei Wu$^1$ and Xi Jin$^1$*}
\newcommand\hpcaaffiliation{$^1$Department of Physics, University of Science and Technology of China, \\
$^2$University of Washington, $^3$Raytron Technology 
}
\newcommand\hpcaemail{wqz1998@mail.ustc.edu.cn, xiaotian.wang@raytrontek.com, jinxi@ustc.edu.cn} 

\author{
  \ifdefined\hpcacameraready
    \IEEEauthorblockN{\hpcaauthors{}}
      \IEEEauthorblockA{
        \hpcaaffiliation{} \\
        \hpcaemail{}
      }
  \else
    \IEEEauthorblockN{\normalsize{HPCA \hpcayear{} Submission
      \textbf{\#\hpcasubmissionnumber{}}} \\
      \IEEEauthorblockA{
        Confidential Draft \\
        Do NOT Distribute!!
      }
    }
  \fi 
}

\fancypagestyle{camerareadyfirstpage}{%
  \fancyhead{}
  
  \fancyhead[C]{
    \ifdefined\aeopen
    \parbox[][12mm][t]{13.5cm}{\hpcayear{} IEEE International Symposium on High-Performance Computer Architecture (HPCA)}    
    \else
      \ifdefined\aereviewed
      \parbox[][12mm][t]{13.5cm}{\hpcayear{} IEEE International Symposium on High-Performance Computer Architecture (HPCA)}
      \else
      \ifdefined\aereproduced
      \parbox[][12mm][t]{13.5cm}{\hpcayear{} IEEE International Symposium on High-Performance Computer Architecture (HPCA)}
      \else
      \parbox[][0mm][t]{13.5cm}{\hpcayear{} IEEE International Symposium on High-Performance Computer Architecture (HPCA)}
    \fi 
    \fi 
    \fi 
    \ifdefined\aeopen 
      \includegraphics[width=12mm,height=12mm]{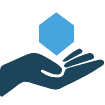}
    \fi 
    \ifdefined\aereviewed
      \includegraphics[width=12mm,height=12mm]{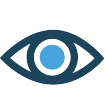}
    \fi 
    \ifdefined\aereproduced
      \includegraphics[width=12mm,height=12mm]{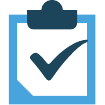}
    \fi
  }
  \fancyfoot[C]{}
}
\begin{document}
\maketitle

%Enables the camera ready header and footer
\ifdefined\hpcacameraready 
  \thispagestyle{camerareadyfirstpage}
  \pagestyle{empty}
\else
  \thispagestyle{plain}
  \pagestyle{plain}
\fi

\newcommand{\hpcaheight}{0mm}
\ifdefined\eaopen
\renewcommand{\hpcaheight}{12mm}
\fi

%%%%%%%%%%%%%%%%%%%%%%%%%%%%%%%%%%%%%%%%
%%%%%%%% -- PAPER CONTENT STARTS -- %%%%%%%%%
% \textcolor{red}{
\begin{abstract}

General matrix-matrix multiplication (GEMM), serving as a cornerstone of AI computations, has positioned tensor processing engines (TPEs) as increasingly critical components within existing GPUs and domain-specific architectures (DSA).
Our analysis identifies that the prevailing architectures primarily focus on dataflow or operand reuse strategies, when considering the combination of matrix multiplication with multiply-accumulator (MAC) itself, it provides greater optimization space for the design of TPEs. This work introduces a novel perspective on matrix multiplication from a hardware standpoint, focusing on the bit-weight dimension of MACs. Through this lens, we propose a finer-grained TPE notation, using matrix triple loops as an example, introducing new methods and ideas for designing and optimizing PE microarchitecture. Based on the new notation and transformations, we propose four optimization techniques that achieve varying degrees of improvement in timing, area, and power consumption. We implement our design in RTL using the SMIC-28nm process. Applying our methods to four classic TPE architectures (include systolic array\cite{jouppi2023tpu}, 3D-Cube\cite{ascend}, multiplier-adder tree\cite{Trapezoid}, and 2D-Matrix\cite{lu2017flexflow}), we achieved area efficiency improvements of 1.27$\times$, 1.28$\times$, 1.56$\times$, and 1.44$\times$, and 1.04$\times$, 1.56$\times$, 1.49$\times$, and 1.20$\times$ for energy efficiency respectively. When applied to a bit-slice architecture, we achieved a 12.10$\times$ improvement in energy efficiency and 2.85$\times$ in area efficiency compared to Laconic \cite{sharify2019laconic}. Our Verilog HDL code, along with timing, area, and power reports for circuit synthesis in URL: \href{https://github.com/wqzustc/High-Performance-Tensor-Processing-Engines}{\color{blue}{https://github.com/wqzustc/High-Performance-Tensor-Processing-Engines}}.

\end{abstract}

\section{Introduction}
In the current wave of technological innovation, artificial intelligence (AI) has become central to modern technological development \cite{gpt}. All these advancements rely on tensor operations, specifically matrix multiplication (MM), which is considered the cornerstone of deep learning and AI. To meet the increasing computational demands of AI, hardware designers are now integrating specialized MM units called tensor processing engines (TPEs) into GPUs \cite{nvidia}, CPUs \cite{Qualcomm,Apple}, and DSAs \cite{googletpu,ipu,sambanova,ascend,mlu}. TPE occupies an important part of the area and power in these chips.

\begin{figure} [htbp]
  \flushleft 
  \includegraphics[scale=0.327]{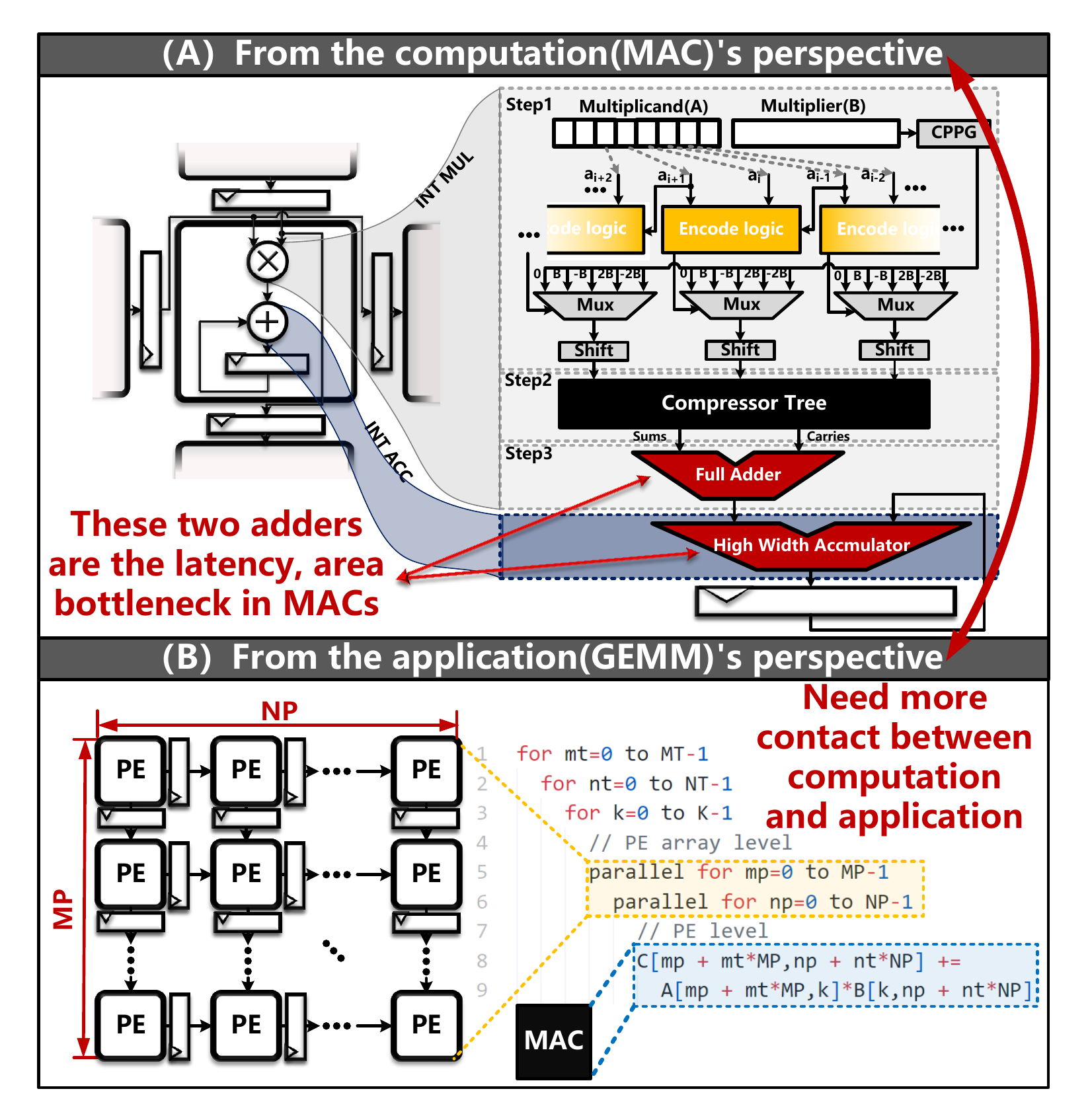}
  \caption{The microarchitecture of INT MAC and MM unit.}
  \label{introduction_1}
\vspace{-0.1cm}
\end{figure}

The MAC, as a fundamental hardware component, has been extensively studied to better exploit MAC and improve computational performance. The research can be divided into macro-architectural level and micro-arithmetic logic level.

On the one hand, architects have designed various architectures for MM based on MAC (Figure \ref{introduction_1}(B), including 2D-matrix \cite{lu2017flexflow}, weight stationary (WS) or output stationary (OS) systolic array \cite{norrie2021design, jouppi2017datacenter, jouppi2023tpu, feldmann2023spatula}, and 3D-Cube \cite{ascend,v100}. These architectures have not only gained widespread application in academia but have also been practically deployed in the industry, becoming foundational in TPE design.

\begin{figure*} [htbp]
  \flushleft 
  \includegraphics[scale=0.191]{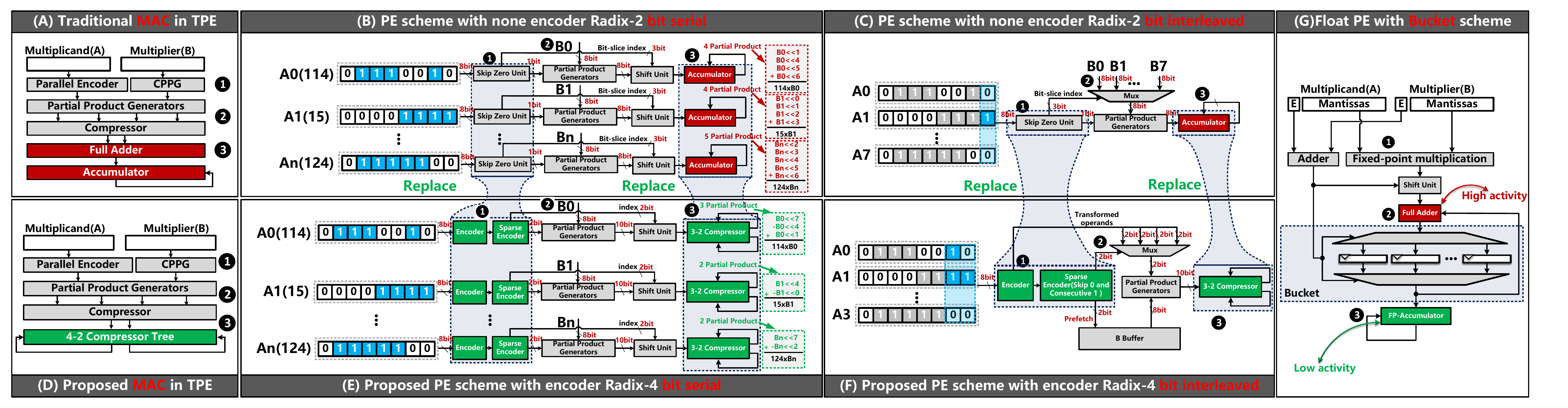}
  \caption{Improvements in microarchitecture compared to other works. (A) Traditional MAC (TPU-Like). (B) and (C) Bit-serial-based computation methods. (D) Optimized MAC. (E) and (F) Optimized bit-serial architectures. (G) Similarities and differences with floating-point optimized schemes. Without showing the DFFs, only Step \ding{184} includes a pipeline register, while the other steps are single-cycle operations.}
  \label{introduction_2}
\vspace{-0.1cm}
\end{figure*}

On the other hand, arithmetic logic units (ALUs) researchers focus on the approach from the perspective of single computational (Figure \ref{introduction_1}(A)), striving to develop higher performance multipliers and adders. Typical designs include array multipliers \cite{huang2005high}, Booth multipliers \cite{kuang2009modified}, Baugh multipliers \cite{sjalander2008high}, carry lookahead adders \cite{cheng2000self}, carry select adders \cite{bedrij1962carry}, carry save adders \cite{gustafsson2004multiplier}, Wallace tree \cite{wallace1964suggestion} and compressor tree\cite{santoro1989spim,veeramachaneni2007novel}.

Architecture design and computational unit design are orthogonal approaches, both of which play a crucial role in enhancing the performance of computer systems. However, current research often focuses solely on one of these two aspects, overlooking the deeper connection between them.

From a comprehensive perspective, traditional MAC (Figure \ref{introduction_2}(A)) is mainly divided into three stages: \ding{182} Encode the multiplicand, and generate partial products (PPs). \ding{183} Compress all PPs to generate the final sum and carry. \ding{184} Obtain the final result through the processing of full adders and accumulators. It has been proved that the internal maximum logic propagation delay ($t_{pd}$) and area of high bit-width accumulation units in MACs have become bottlenecks \cite{blumenfeld2024towards} for performance and efficiency (Figure \ref{introduction_1}(A) step \ding{184}). One solution (Figure \ref{introduction_2}(G)) proposed by Bucket Getter \cite{BucketGetter} allows a large number of floating-point additions be converted to fixed-point additions during the reduction (accumulation) phase (Figure \ref{introduction_2}(G)\ding{183}). It significantly reduces the dependency on floating-point accumulators and lowers the activity. This method reduces the power consumption of the floating-point accumulators (Figure \ref{introduction_2}(G)\ding{184}) and further improves energy efficiency within the process element (PE).  \textbf{Q\textcolor{blue}{I}:\label{q1} However, the issue of high-bit-width fixed-point accumulation bottleneck ($t_{pd}$ and area) remains unresolved in this research.}

Differing from the MAC in parallel, the researchers have proposed bit-slice-based multiplication methods to replace MAC. These methods main include the Radix-2 bit-serial \cite{judd2016stripes,delmas2019bit,wang2024bsvit,pan2023bitset}, Radix-2 bit-interleaved \cite{lu2021distilling,shi2024bitwave,yang2021fusekna,albericio2016bitpragmaticdeepneuralnetwork}, higher width bit-slice \cite{im2023sibia} and Radix-4 based slice computation \cite{sharify2019laconic,im2024lutein,li2022ristretto,grimm2024training}. The Radix-2 bit-serial computation (shown in Figure \ref{introduction_2}(B)) relies on the sparsity of the multiplicand A and consists of three major steps. Step \ding{182} is to extract the indices of non-zero bit slices of A and skip zero elements. Step \ding{183} uses these indices and the multiplier B to generate PPs. Step \ding{184}  is to shift and accumulate these PPs according to their corresponding bit-weight. The computation speed of this method mainly depends on the number of PPs, or the number of non-zero bit slices in A. The Radix-2 bit-interleaved computation method (Figure \ref{introduction_2}(C)) processes multiple data simultaneously. These data from different operands share the same bit-weight, thus eliminating the need for shift operations. Step \ding{184} usually consists of an adder-tree or a high bit-width accumulator. The radix-2 multiplication calculation method can maintain high bit sparsity, but it has the disadvantage of requiring a large number of PPs to be accumulated.

Despite the achievements in improving computational efficiency, these bit-slice-based methods still have room for improvement. \textbf{Q\textcolor{blue}{II}:\label{q2} Firstly, these methods cannot effectively skip consecutive ``1" bit-slice, which may affect computational efficiency in certain cases (under the two's complement representation of a batch of normally distributed data, the number of bit-slices with ``1"s in negative numbers is typically greater than those with ``0"s). Secondly, the accumulation can still become a performance bottleneck.}

In encoder-based TPE design schemes, such as Pragmatic \cite{albericio2016bitpragmaticdeepneuralnetwork} and Laconic \cite{sharify2019laconic}, encoder-based strategies are used to generate partial products. Both leverage the bit sparsity of encoded data to accelerate DNNs. However, the fundamental principles underlying the use of encoders for sparsity acceleration have not been thoroughly explored.

To address the aforementioned issues, this study proposes a PE microarchitecture design method tailored for specific tasks. We propose an analytical model of the MAC based on a compute-centric notation. This model assists us in better integrating the concepts presented in Figure \ref{introduction_1}(A) and (B) from a deeper and more intuitive perspective, as well as mapping the hardware components within the MAC to corresponding primitives, such as parallel encoder, candidate partial product generator (CPPG), partial product generator, shifter, compressor, full adder, and high bit-width accumulator. Through this transformation, we uncover the implicit parallel dimension within the MAC units in the new notation. This dimension was overlooked in the previous TPE design.

\begin{figure} [htbp]
  \flushleft 
  \includegraphics[scale=0.28]{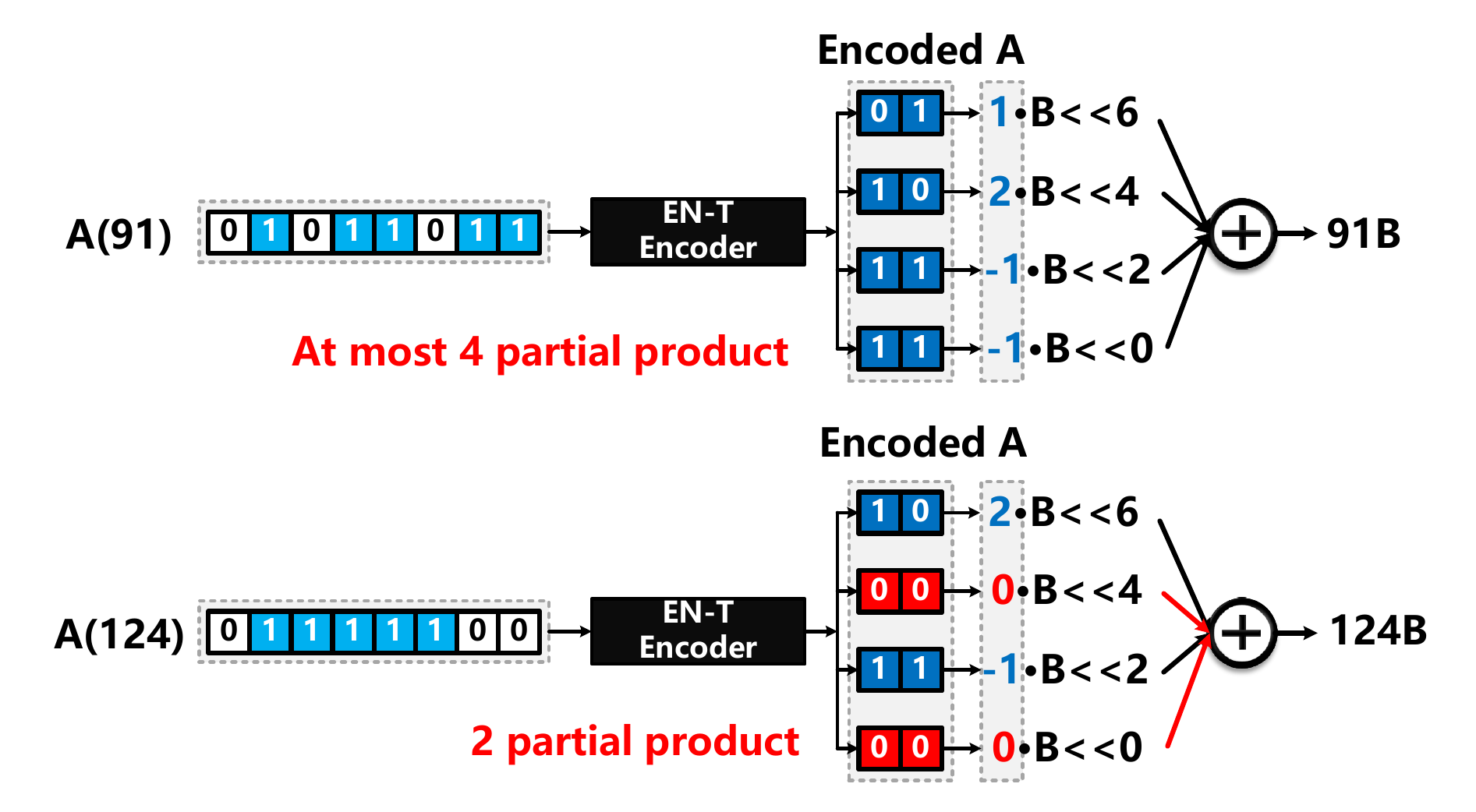}
  \caption{Example of multiplication based on encoding.}
  \label{back_1}
\vspace{0cm}
\end{figure}

By the transformation under the new notation, we design several optimized PE microarchitectures for Q\hyperref[q1]{\textcolor{blue}{I}} and Q\hyperref[q2]{\textcolor{blue}{II}} (Figure \ref{introduction_2}(D)(E) and (F)). In matrix multiplication, MAC usually involves vector reduction in the time dimension. Therefore, before the reduction process is completed, we use compressors \cite{veeramachaneni2007novel} to perform accumulation (store the sums and carries in DFFs (Data Flip-Flops)) to replace the full adder in step \ding{184} of Figure \ref{introduction_2}(A) and (D). Experiments show that even under high bit-width ($\geq$32bit) accumulation conditions, our design can reduce the $t_{pd}$ of traditional MAC by half in Q\hyperref[q1]{\textcolor{blue}{I}}. This is because the delay of the half-adder is not dependent on the operand bit-width (shown in Table \ref{csa}).

To solve Q\hyperref[q2]{\textcolor{blue}{II}} in bit-serial computation, we employ encoding and half-accumulation strategies (Figure \ref{introduction_2}(E)). In step \ding{182}, encoders are used to replace the original Skip Zero Unit. The encoding can utilize modified Booth encoding (MBE) \cite{kuang2009modified} or EN-T \cite{wu2024entensorcore} encoding. Subsequently, \textbf{sparse encoding is performed on the encoded numbers}, which differs from other bit-slice-based calculations. While other schemes perform sparse encoding on the ``0" bit-slice of the multiplicand, we conduct sparse encoding on the encoded number. This is because the encoded number is directly utilized to generate the PPs (Figure \ref{introduction_1}(A) step \ding{182}).

For comparison purposes, two computational examples are presented in Figure \ref{introduction_2}(B) and (E), where 114, 15, and 124 are multiplied by three multipliers $B_0, B_1$ and $B_n$. Bit-serial computation generates 4, 4, and 5 PPs respectively, requiring corresponding cycles to complete accumulation. In contrast, our proposed method only requires 3, 2, and 2 PPs for these operands. This allows for skipping both zeros and consecutive ``1" bit-slices in the multiplicand.

The second improvement involves the use of compressors to replace accumulators in step \ding{184} of Figure \ref{introduction_2}(B)(E). This is aimed at optimizing area and timing. In terms of optimization for bit-interleaved methods (Figure \ref{introduction_2}(F)), this paper employs sparse encoding for the encoded bits of A. \textbf{The sparsely encoded indices are used to select the encoded bits, while also utilizing the indices to prefetch B}. Subsequently, PPs are generated by using the encoded bit segments of non-zero multiplicand A and multiplier B. These PPs are then accumulated using 3-2 compressor to optimize computational efficiency.

In summary, our core contributions are as follows:

\begin{enumerate}
\item We propose a finer-grained TPE notation and introduce new methods and ideas for designing and optimizing PE microarchitecture in specific applications.
\item Compared to Pragmatic\cite{albericio2016bitpragmaticdeepneuralnetwork} and Laconic\cite{sharify2019laconic}, we provide a more systematic explanation of the fundamental reasons for bit-sparse acceleration and design a more efficient PE micro-architecture suitable for bit-serial processing, characterized by low area and high frequency. Additionally, we discuss the comparison of other encoding methods for bit-sparse acceleration of the multiplicand.
\item Based on the new notation and transformations, we propose four optimization methods and we implement our design in RTL using the 28nm process. Applying our methods to four classic TPE architectures (include systolic array\cite{jouppi2023tpu}, 3D-Cube\cite{ascend}, multiplier-adder tree\cite{Trapezoid}, and 2D-Matrix\cite{lu2017flexflow}), we achieved area efficiency improvements of 1.27$\times$, 1.28$\times$, 1.56$\times$, and 1.44$\times$, and 1.04$\times$, 1.56$\times$, 1.49$\times$, and 1.20$\times$ for energy efficiency respectively. When applied to a bit-slice architecture, we achieved a 12.10$\times$ improvement in energy efficiency and 2.85$\times$ in area efficiency compared to Laconic \cite{sharify2019laconic}.

\end{enumerate}

\section{Background and Motivation}

\subsection{High Width Accumulator Represents the Most Challenge}
For AI DSA, the performance of the TPE is key to ensuring DNN throughput. For example, in TPU\cite{jouppi2017datacenter}, the systolic array and accumulators occupy 36\% of the die area and consume 52\% of the on-chip power; for Bitwave\cite{shi2024bitwave}, TPE accounts for 26.8\% of the area and 62.5\% of the power consumption; for LUTein\cite{im2024lutein}, TPE takes up 27.7\% of the area and 51.6\% of the on-chip power; for Bucket\cite{BucketGetter}, TPE occupies 59.5\% of the area and 33.7\% of the on-chip power. Therefore, improving the performance of TPE within the limited on-chip silicon area is crucial for AI DSA.

\begin{table}[]\centering
\begin{tabular}{ccccc}
\midrule\midrule
\textbf{Unit}      & \textbf{Bit} & \textbf{Area($um^{2}$)}    & \textbf{Delay($ns$)}  & \textbf{TOP($uW$)}\\ \midrule
\multirow{4}{*}{MAC}        & 20    & 179.30  & 1.56  & 27.1 \\
                            & 24    & 192.65  & 1.67  & 29.2\\
                            & 28    & 206.01  & 1.84  & 31.4\\
                            & 32    & 238.51  & 1.97  & 36.3\\ \midrule
4-2 Compressor Tree                     & 14    & 55.92   & 0.31 & 8.5  \\ \midrule
Full Adder                  & 14    & 51.32   & 0.34  & 7.7 \\ \midrule
\multirow{4}{*}{Accmulator} & 20    & 57.32   & 0.80  & 8.6\\
                            & 24    & 62.43   & 0.90  & 9.4\\
                            & 28    & 82.78   & 0.99  & 12.3\\
                            & 32    & 95.13   & 1.13  & 14.3\\ \midrule\midrule
\end{tabular}
\caption{The main component decomposition in INT8 MAC (tested on SMIC 28nm with a 2ns clock constraint).}
\label{mac}
\end{table}

For TPE, the area and $t_{pd}$ of the MACs are the primary performance bottlenecks. Within the MAC, the compressor tree and the full adder within the multiplier are affected by the multiplication bit-width in terms of logical delay and area, while the bit-width of the accumulator is only related to the number of accumulations required. Consequently, as the bit-width of the accumulator increases, it gradually becomes a limiting factor for the MAC's frequency. According to the circuit synthesis report listed in Table \ref{mac}, it's observed that with the bit-width of the accumulator increases in MAC, the predominant factors constraining the performance progressively transition to the area and delay of the accumulator. For instance, in a 32-bit accumulation, the logical area occupied by the full adders and accumulator accounts for 61.4\%, and the logic delay is as high as 74.6\%, severely restricting the frequency of the MACs.

\subsection{Fine-grained Description of the TPE Microarchitecture}

The RTL-based description of the TPE microarchitecture is overly detailed for designers to understand the acceleration mechanism at the algorithm level, while the hardware block diagram representation is too abstract for the underlying implementation level. Using a notation between the hardware block diagram and RTL can help designers understand the acceleration mechanism at both levels. 
However, existing design space representations \cite{maestro,interstellar,tileflow,zigzag,timeloop} focus on architecture with MAC as the basic unit and don't explicitly represent the reduction logic (adder-trees) brought by spatial unrolling and the reduction logic in PEs constitutes a significant portion of the PE area and serves as a critical factor that affects timing. These limitations make it difficult to capture acceleration opportunities at the PE microarchitecture level. Therefore, there is a need to develop a more comprehensive notation for TPE to capture data flow and operational specifics in detail, enabling further exploration of hardware architecture optimization methods under specific application conditions.

\subsection{Sparsity Acceleration Based on the Encoding Principle}
\label{sec:motivation_sparsity}

In previous research based on bit-serial methods \cite{lu2021distilling,shi2024bitwave,im2023sibia}, the sparsity of bit-slice was often used to discuss potential speed improvements while overlooking the number of partial products (NumPPs) in multiplication. However, the NumPPs directly influence hardware delay and area for parallel multiplication, as well as the number of cycles needed for serial multiplication.

\begin{table}[]
\centering

\begin{tabular}{cccccc}
\midrule\midrule
\textbf{NumPPs} & \textbf{4}                                                      & \textbf{3}                                                      & \textbf{2}                                                      & \textbf{1}                                                     & \textbf{0}                                                    \\ \midrule
MBE\cite{8_farooqui1998general}    & \begin{tabular}[c]{@{}c@{}}81\\ \textbf{\color{red}{(31.6\%)}}\end{tabular} & \begin{tabular}[c]{@{}c@{}}108\\ \textbf{\color{cyan}{(42.2\%)}}\end{tabular} & \begin{tabular}[c]{@{}c@{}}54\\ \textbf{\color{cyan}{(21.1\%)}}\end{tabular} & \begin{tabular}[c]{@{}c@{}}12\\ \textbf{\color{cyan}{(4.7\%)}}\end{tabular} & \begin{tabular}[c]{@{}c@{}}1\\  \textbf{\color{cyan}{(0.4\%)}}\end{tabular} \\
EN-T\cite{wu2024entensorcore}   & \begin{tabular}[c]{@{}c@{}}72\\ \textbf{\color{red}{(28.1\%)}}\end{tabular} & \begin{tabular}[c]{@{}c@{}}108\\ \textbf{\color{cyan}{(42.2\%)}}\end{tabular} & \begin{tabular}[c]{@{}c@{}}60\\  \textbf{\color{cyan}{(23.4\%)}}\end{tabular} & \begin{tabular}[c]{@{}c@{}}15\\ \textbf{\color{cyan}{(5.9\%)}}\end{tabular} & \begin{tabular}[c]{@{}c@{}}1\\  \textbf{\color{cyan}{(0.4\%)}}\end{tabular} \\ \midrule\midrule

\textbf{NumPPs} & \textbf{\{8,7\}}                                                      & \textbf{\{6,5\}}
& \textbf{4}         & \textbf{\{3,2\}}     & \textbf{\{1,0\}}  
 \\ \midrule
 bit-serial & \begin{tabular}[c]{@{}c@{}}9\\ \textbf{\color{red}{(3.5\%)}}\end{tabular} & \begin{tabular}[c]{@{}c@{}}84\\ \textbf{\color{red}{(32.8\%)}}\end{tabular} & \begin{tabular}[c]{@{}c@{}}70\\ \textbf{\color{red}{(27.3\%)}}\end{tabular} & \begin{tabular}[c]{@{}c@{}}84\\  \textbf{\color{cyan}{(32.8\%)}}\end{tabular} & \begin{tabular}[c]{@{}c@{}}9\\  \textbf{\color{cyan}{(3.5\%)}}\end{tabular} \\
\midrule\midrule
\end{tabular}
\caption{The number of partial products (NumPPs) under different encoding within the range of INT8 ($-128 \sim 127$).}
\label{table1}
\vspace{0.2cm}
\end{table}

\begin{table}[]\centering
\begin{threeparttable}  
\begin{tabular}{cccccc}
\midrule\midrule
\textbf{Distribution} & \textbf{$\mathcal{N}(0,0.5)$} & \textbf{$\mathcal{N}(0,1.0)$}  & \textbf{$\mathcal{N}(0,2.5)$} & \textbf{$\mathcal{N}(0,5.0)$} \\ \midrule
\textbf{\color{cyan}{EN-T}}         & \textbf{\color{cyan}{2.27}}      & \textbf{\color{cyan}{2.22}}  & \textbf{\color{cyan}{2.26}}  & \textbf{\color{cyan}{2.23}}              \\ 
MBE          & 2.46      & 2.41  & 2.45  & 2.42        \\
bit-serial(M)\tnote{\ding{182}}  & 3.52      & 3.52  & 3.52   & 3.53\\
bit-serial(C)\tnote{\ding{183}}    & 3.99      & 3.98  & 3.98   & 3.98          \\\midrule\midrule
\end{tabular}
         \begin{tablenotes} 
        \footnotesize            
        \item[\ding{182}] Operand with complement representation.         
        \item[\ding{183}] Operand with sign-magnitude representation.      
      \end{tablenotes}            
 \end{threeparttable}
\caption{The average NumPPs of each multiplicand in different encoding based on the normal distribution matrix.}
\vspace{0.1cm}
\label{table2}
\end{table}

\begin{table*}[]
\centering
\begin{tabular}{ll}
\midrule\midrule
\textbf{PRIMITIVE}                    & \textbf{DESCRIPTION}                                                                                                                                                                                                                                                                                  \\ \midrule
% \textbf{$half\_reduce(I_1, I_2,…,I_n)$} & Half compress tree with $n$ inputs, inputs $I_1$ to $I_n$, and outputs the $sum$ and the $carry$.                                                                                                                                                                                                             \\  \midrule[0.2pt]
\textbf{$half\_reduce(I_1, I_2,\dots,I_n)$} & Compressor tree, with $n$ inputs ($I_1$ to $I_n$) and $2$ outputs (sum and carry).                                                                                                                                                                                                             \\  \midrule[0.2pt]
\textbf{$add(I_1, I_2)$}                  & Full adder, with $2$ input and $1$ output.                                                                                                                                                                                                                                                                            \\   \midrule[0.2pt]
\textbf{$accumulate(I)$}              & \begin{tabular}[c]{@{}l@{}}Accumulator, unlike the full adder, has inputs that depend on the output of the previous cycle.\end{tabular} \\ \midrule[0.2pt]
% \textbf{$encode(I, i)$}                     & \begin{tabular}[c]{@{}c@{}}The input of the encoder is $I$, and $i$ bit weights, related to the encoding algorithm, \\ such as in MBE, $[2i-1:2i+1]$ three bits are selected as input,  and three signals NEG, SE, CE are outputted\end{tabular}                                         \\ \midrule[0.2pt]
\textbf{$encode(I, i)$}                     & \begin{tabular}[c]{@{}l@{}}Encoder, outputs candidate PPs selection signal. $I$ represents bit-slice of multiplicand, and $i$ is the $i$-th bit weight. \\
In MBE, $i \in [0,3]$ and $I$ consists of $[2i-1:2i+1]$ slice from multiplicand. \end{tabular}                                         \\ \midrule[0.2pt]
% \textbf{$map(I, sel)$}                  & \begin{tabular}[c]{@{}c@{}}Partial product generator as well as selector. It generates the partial products \\ required for encoding based on the input $I$ and selects the corresponding partial products based on the $sel$ signal.\end{tabular}                                          \\  \midrule[0.2pt]
\textbf{$map(I, sel)$}                  & \begin{tabular}[c]{@{}c@{}} CPPG and Mux, $``map"$ generates the PPs based on the input $I$ through selects the corresponding PPs based on the $sel$.\end{tabular}                                          \\  \midrule[0.2pt]
% \textbf{$shift(I, i)$}                  & \begin{tabular}[c]{@{}c@{}}Shifter, related to the encoding algorithm. Shift function \\indicates the number of bits to be shifted at $i$ bit weights, which is \textless{}\textless{}$2i$ in MBE\end{tabular}                                                                           \\ \midrule\midrule
\textbf{$shift(I, i)$}                  & \begin{tabular}[c]{@{}c@{}}Shifter, $I$ denotes the data to be shifted, while $i$ is the configuration. $I$ will be shifted to the left by $2i$ bits in MBE.\end{tabular}                                                                           \\ \midrule\midrule
\end{tabular}
\caption{Components described by hardware primitives.}
\label{table4}
\vspace{-0.2cm}
\end{table*}

\begin{figure*} [htbp]
  \flushleft 
  \includegraphics[scale=0.274]{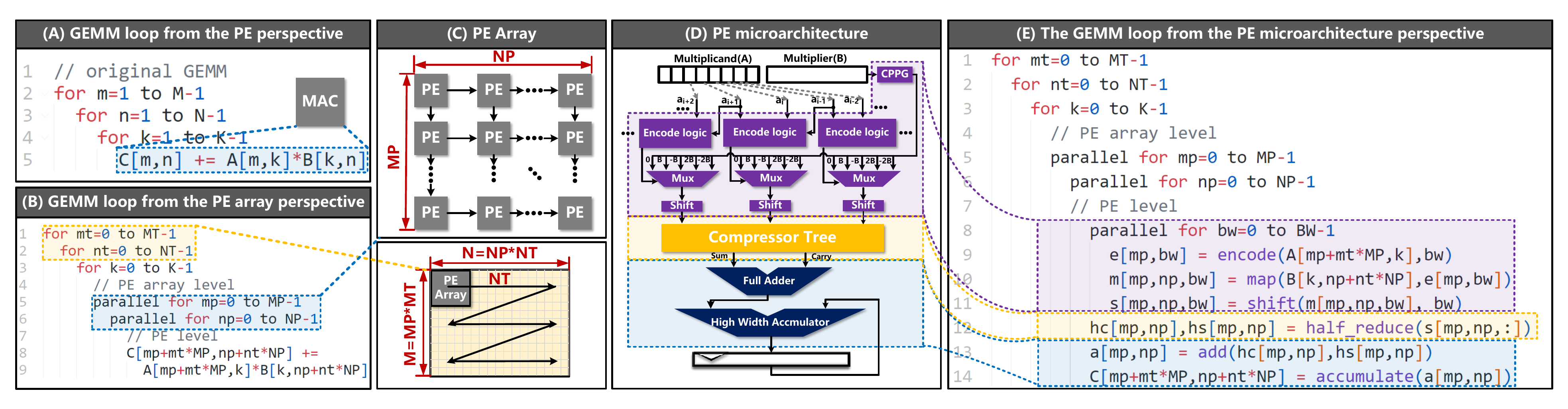}
  \caption{The GEMM loop from the PE microarchitecture perspective.}
  \label{method_1}
\vspace{-0.1cm}
\end{figure*}

For a Radix-4 parallel multiplier, an $n$ bit multiplicand processed by an encoder (MBE \cite{8_farooqui1998general} or EN-T \cite{wu2024entensorcore}) will produce $\frac{n}{2}$ PPs. Taking Radix-4 EN-T as an example (Figure \ref{back_1}(A)), for INT8 multiplication, the multiplicand A (in two's complement) generates four 2-bit encoded numbers after passing through the encoder. For instance, with 91, the encoded numbers are \{1, 2, -1, -1\}, corresponding to PPs coefficients with bit-weight \{$2^6, 2^4, 2^2, 2^0$\}. Therefore, multiplying the multiplier B by 91 can be expressed as four PPs: 91B = (B$<<$6) + (2B$<<$4) + (-B$<<$2) + (-B). The candidate PPs only need to compute \{-2B, -B, 0, B, 2B\} in MBE for selection by the encoded numbers, and the shifter is responsible for shifting the generated PPs by the corresponding bits weight.

However, not all numbers will produce 4 non-zero PPs (Figure \ref{back_1}(B) 124 can be encoded as \{2, 0, -1, 0\}, so 124B = (2B$<<$6) + (-B$<<$2)). We counted the NumPPs generated by the two Radix-4 encoders and Radix-2 bit-serial within the INT8 range (Table \ref{table1}). Under MBE, 175 ((108 + 54 + 12 + 1)/256 $\approx$ \textbf{68.4\%}) numbers generate 3 or fewer non-zero PPs during multiplication. Under EN-T, 184 ((108 + 60 + 15 + 1)/256 $\approx$ \textbf{71.9\%}) numbers generate 3 or fewer non-zero PPs. Similarly, Radix-2 bit-serial complement multiplication can be viewed as generating PPs without encoding. Only 93 ((84 + 9)/256 $\approx$ \textbf{36.3\%}) numbers generate 3 or fewer non-zero PPs during multiplication.

To assess the overall operational cost of batch data, we use the average NumPPs as a metric. Fewer PPs lead to faster computation and lower power consumption. In Table \ref{table2} (matrices size $1024 \times 1024$), the average NumPPs for EN-T and MBE range from 2.22 to 2.45. Therefore, for large-scale matrix multiplication, we break down the vector dot product operation into two key steps: \ding{182} Generating non-zero partial products. \ding{183} Reduction of partial products.  For parallel MAC, the multiplicand is encoded during the computation of the vector dot product, and the partial products are expanded spatially as an implicit dimension for parallel processing and reduction. This process ignores the scenario where some of the generated partial products are zero. Thus, by decomposing the multiplication operation into sequential partial product reduction combined with a non-zero partial product generator, the number of operations in matrix multiplication can be significantly reduced.

\section{Proposed Methdology}

In this study, we employ a compute-centric notation that closely resembles software pseudocode to improve comprehensibility. This notation is utilized to depict the microarchitecture of TPEs by incorporating the bit-weight dimension (referred to as $BW$) of MACs. Subsequently, we will examine the new hardware primitives introduced by the $BW$ and provide an example of TPEs utilizing a 2D-OS dataflow within the notation. Our goal is to offer a clear and professional perspective on how the $BW$ of MACs impacts the performance of TPEs.

\subsection{BW Dimension and New Hardware Primitives}

In a multiplier, the calculation process can be visualized as a multiplicand expanded into multiple sub-operands, which are then multiplied in parallel with another operand (resulting in the PPs of different bit-weights), and finally reducing all PPs to obtain the result. 
% \modified{The number of sub-operands is the size of the $BW$ dimension.}
This can be expressed as follows:

\begin{equation}
    C=A\times B=\sum_{bw=0}^{BW-1}SubA_{bw} \times B.
    \label{eq:multiply}
\end{equation}
% where $BW$ and $SubA_{bw}$ are specific to the encoder being used. 
It should be noted that the size of $BW$ and the form of $SubA_{bw}$ are related to specific encoding methods. In this paper, we focus on the acceleration opportunities brought by the $BW$ dimension, rather than the design of specific encoding methods. Here, we only use two examples to illustrate that Eq. \eqref{eq:multiply}  can broadly represent the multipliers.
For an 8-bit MBE \cite{863039}, $SubA_{bw}$ and $BW$ are as follows:
\begin{equation}
    SubA_{bw} = (-2a_{2bw+1}+a_{2bw}+a_{2bw-1})2^{2bw}, BW=4,
    \label{eq:mbe}
\end{equation}
where $a_{2bw}$ represents the $2bw$-th bit of $A$. For an 8-bit complement bit-serial method \cite{im2023sibia}, $SubA_{bw}$ as follows:
\begin{equation}
\
SubA_{bw} = \begin{cases}
a_{bw}2^{bw}, &if \  bw\neq BW-1\\
-a_{bw}2^{bw}, &if \  bw=BW-1
\end{cases}, BW=8.
\label{eq:bs}
\end{equation}

Based on Eq. \eqref{eq:multiply}, the multiplier exposes its implicit dimension $BW$, which represents the number of sub-operands of $A$. Based on Eq. \eqref{eq:mbe} and \eqref{eq:bs}, each sub-operand can be represented as the encoding of a bit-slice multiplied by a weight. Therefore, we call this hidden dimension the bit-weight dimension. In matrix multiplication, we apply Eq. \eqref{eq:multiply} to obtain the following form:

\begin{equation}
   C_{m,n}=\sum_{k=0}^{K-1} A_{m,k}B_{k,n}=\sum_{k=0}^{K-1}\sum_{bw=0}^{BW-1}SubA_{m,k,bw}B_{k,n}.
\label{eq5}
\end{equation}

\begin{figure*} [htbp]
  \flushleft 
  \includegraphics[scale=0.26]{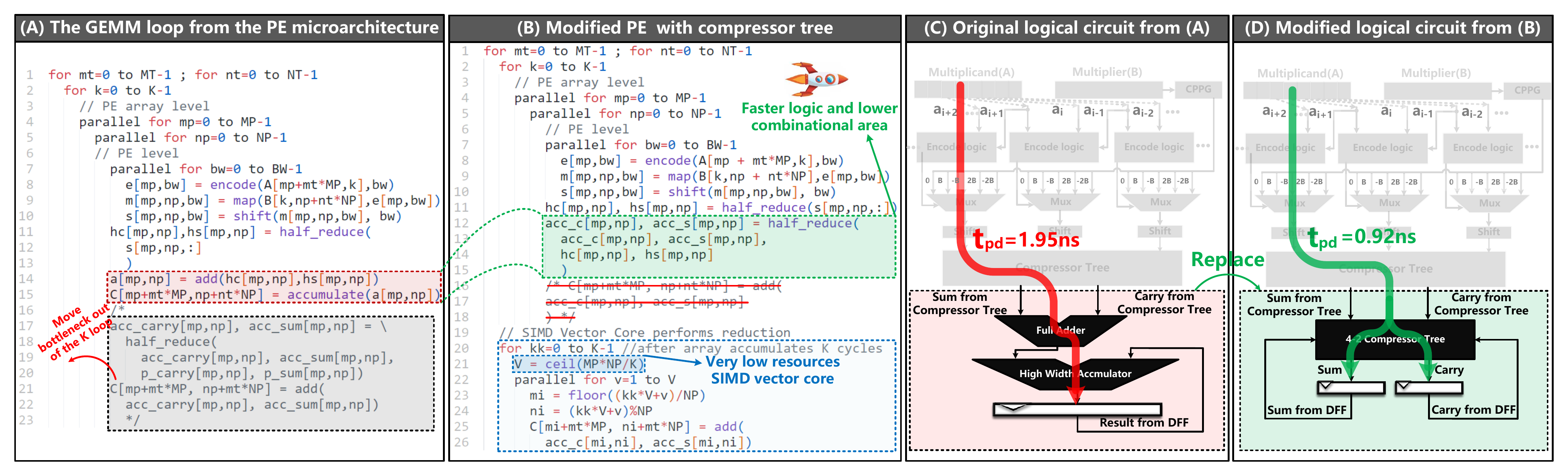}
  \caption{The proposed optimization architecture 1 (OPT1).}
  \label{opt1}
\vspace{-0.1cm}
\end{figure*}

The primary objective of this paper is to utilize the microarchitectural hardware diversity uncovered by $BW$ in order to investigate potential optimization opportunities for TPEs. To precisely demonstrate the impact of the new dimensional transformation on hardware design, we introduced new primitives and explicitly represented these components in the notation. The primitives are shown in Table \ref{table4}.

In the process of multiplication, there are four key components involved in generating the PPs: encoders, CPPGs, multiplexers, and shifters. We utilize the terms \textbf{``encode"}, \textbf{``map"} and \textbf{``shift"} to denote these components. The reduction logic in the MAC, including the compressor tree, full adder, and accumulator, not only takes up a significant area within the PE but also has a crucial impact on timing. In light of this, we have introduced \textbf{``half\_reduce"}, \textbf{``add"}, and \textbf{``accumulate"} to explicitly represent the reduction logic in the notation.

In the following sections, we will investigate how $BW$ transformation impacts TPE microarchitecture by analyzing these components and their relevance. Based on this analysis, we will develop more efficient parallel hardware.

\subsection{Matrix Multiplication from a Microarchitecture Perspective}
\label{secb}

Starting with the traditional triple-nested loop of MM (Figure \ref{method_1}(A)) and the compute-centric notation form (Figure \ref{method_1}(B)), we propose Figure \ref{method_1}(E) as the new notation for the TPE by introducing the $BW$ and new computational primitives.

As illustrated in Figure \ref{method_1}(B), the dimensions $M$ and $N$ are split into 4 sub-dimensions. The $M_T$ and $N_T$ are the temporal sub-dimensions of $M$ and $N$, and the suffix or subscript ``T" refers to \underline{t}emporal dimension. The data is iterated in the zigzag form, and $N_P \times M_P$ loop instances are processed and iterated once at each step within the PE array. The $M_P$ and $N_P$ are the spatial sub-dimensions, while the suffix or subscript ``P" refers to s\underline{p}atial unrolling dimensions. The ``parallel" in pseudo-code means this dimension is mapped to PE array.

Combined with the PE microarchitecture in Figure \ref{method_1}(D), the MAC micro-operation (Figure \ref{method_1}(E)) using primitives from Table \ref{table4}. The \textbf{``encode"} generates the select signal for the Mux, while the \textbf{``map"} produces candidate PPs for the Mux inputs and selects the final PPs. The following equation was derived from these basic primitives:

\begin{small}
\begin{equation}
    \begin{aligned}
&C_{m,n} =\sum_{k=0}^{K-1}\sum_{bw=0}^{BW-1} map(B_{k,n},encode(A_{m,k,bw}))shift(bw)\\
&=\sum_{bw=0}^{BW-1}shift(bw)\sum_{k=0}^{K-1} map(B_{k,n},encode(A_{m,k,bw})).
    \label{eq6}
    \end{aligned}
\end{equation}
 \end{small}

It is obvious that the \textbf{``shift"} is independent to $N$ and relevant to $K$, $M$ and  $BW$ in Eq. (\ref{eq6}), so that \textbf{``shift"} can be decoupled from \textbf{``encode"} and \textbf{``map"}, and be outer level of the $K$ dimension. The movement of \textbf{``shift"} helps to reduce the number of the \textbf{``shift"} in the array. This inspires us to change our position in notation to explore new architectural designs.

In Figure \ref{method_1}(D), the multiplexer outputs one of the candidate PPs based on the select signals. If we represent the select signals as a one-hot vector, then the selection can be viewed as a dot product of the candidate PPs and select vector. Eq. (\ref{eq6}) can be decomposed as follows:

\begin{equation}
% \mathcolorbox{BurntOrange}{
% C_{m,n}=\sum_{bw=1}^{BW}shift(i)\sum_{k=1}^{K}\overrightarrow{enc_{m,k,bw}} \cdot \overrightarrow{prod_{k,n}}
C_{m,n}=\sum_{bw=0}^{BW-1}shift(i)\sum_{k=0}^{K-1}\overrightarrow{enc_{m,k,bw}} \Diamond \overrightarrow{prod_{k,n}},
% }
    \label{eq7}
\end{equation}
where the symbol $\Diamond$ refers to the selection operation. It is a non-commutative operation, as the inputs and select signal of the multiplexer cannot be reversed. The \textbf{``encode"} is independent of $N$ and can be placed outer of the $N$ dimension. The \textbf{``map"} contains the selection operation for the multiplexer instance, so it can only be located in the innermost loop.

% In the preceding analysis, we examine the legality of \textbf{position} and nesting levels in component transformation. 
% The following section will demonstrate the enhancements achieved by altering the \textbf{order} of components.

%  Analyzing the utilization of different components within the PE and finding suitable strategies and methods to optimize the circuit structure can be achieved by uncovering the $BW$ dimension behind the MAC. In the next section, we will optimize the PE microarchitecture using these primitives step by step and demonstrate the entire optimization process.

Other notations derived from Einsum notation focus more on data reuse above the MAC level.
% , and do not change the number of MACs under the same ideal throughput. 
The notation we proposed can represent the hardware implementation of MAC in a fine-grained manner, which is able to represent bit-slice accelerators and allows for the representation of intermediate signal reuse within MACs. 
Based on the preceding analysis of the legality of component positions and nested levels, we can change the position and order of components. Intuitively, changing the nested levels of components can change the number of components, while changing the order can change the critical path of MAC, thereby bringing a new design space dimension to TPE.
% \modified{At the same time, exposing the sub-operands allows us to take advantage of the acceleration opportunity of encoding sparsity discussed in the Sec \ref{sec:motivation_sparsity}.}
Just like the skip-zero in a bit-serial multiplier, under our notation, we can convert the $BW$ dimension to the temporal dimension to skip zero partial products and utilize the sparsity discussed in Sec. \ref{sec:motivation_sparsity}.

In the next section, we will optimize the PE microarchitecture using these primitives step by step and demonstrate the entire optimization process.

\section{Proposed Architecture}

This section delves into optimizations based on the new notation to uncover the potential for latency or area improvements. Conventional design space exploration methods mainly focus on loop transformations and changing spatial mapping dimensions. In contrast to previous studies, our study provides a more detailed analysis of MACs and proposes four orthogonal optimization techniques aimed at enhancing TPE performance within the current notation framework.

\subsection{Half Compress Accumulation Reduction (OPT1)}
\label{secopt1}

In traditional MAC-based TPE (Figure \ref{opt1}(A)), the \textbf{``accumulate"} follows the \textbf{``add"} because the compiler needs to keep the multiplier atomic. Accumulating the output of a full adder is common but costly due to high bit-width results. Fortunately, from a MAC's perspective, it is possible to reverse the order of reduction (\textbf{``accumulate"}) and \textbf{``add"}. This means replacing codes in the red box (line 14 $\sim$ line 15) with those within the gray box (line 16 $\sim$ line 23) in Figure \ref{opt1}(A) keeps the result correct. The reorder results in a faster and smaller logic circuit for the reduction: the accumulation in the compressor tree.

It is essential to note that the \textbf{``add"} depends solely on the accumulated acc\_c and acc\_s (Figure \ref{opt1}(A) line 17). Therefore, the result of the \textbf{``add"} is not needed until the final iteration of the $K$ dimension, when the accumulation of acc\_c and acc\_s is not completed, the computation of the \textbf{``add"} is redundant.

Inspired by the above, we propose an optimization strategy illustrated in Figure \ref{opt1}(B). This strategy uses the half-add operation during the reduction process of $K$ dimension, \textbf{which ensures that the logic delay is independent of the cumulative bit-width}, reducing the need for full adders and accumulators. With only one valid output generated within $K$ cycles per PE, fewer \textbf{``add"} operations are needed to merge the acc\_s and acc\_c at the same level of $K$ dimension. The external full adders (typically a SIMD vector core) outside the PE array handle these \textbf{``add"} operations and work with TPE in parallel. Since the SIMD vector core only accesses the data for every $K$ cycle, hence, fewer hardware resources ($\left\lceil (M_P N_P/K) \right\rceil $) are required to accomplish these tasks.

\begin{table}[]\centering

\begin{tabular}{cccc}
\midrule\midrule
\textbf{Component}                       & \textbf{Width} & \textbf{Area($um^{2}$)}    & \textbf{Delay($ns$)} \\ \midrule
\multirow{6}{*}{4-2 Compressor Tree} & 14    & 52.92   & 0.31  \\
                                   & 16    & 60.98   & 0.32  \\
                                   & 20    & 77.11  & 0.32  \\
                                   & 24    & 93.99  & 0.32  \\
                                   & 28    & 110.12 & 0.32  \\
                                   & 32    & 126.25  & 0.32  \\ \midrule\midrule
\end{tabular}
\caption{Timing and area of the compressor on SMIC 28nm.}
\label{csa}
\end{table}

The hardware architecture of the original PE is illustrated in Figure \ref{opt1}(C), while the proposed in Figure \ref{opt1}(D). This enhancement involves replacing the full adder and accumulator, which currently account for a significant portion of the critical path delay ($t_{pd}$) and area within the PE, with a single 4-2 compressor tree. With a clock constraint of 2ns, we are able to reduce the $t_{pd}$ from 1.95ns to 0.92ns (for INT8 multiplication and INT32 accumulation synthesis on SMIC-28nm), and easily achieve a clock frequency exceeding 1GHz. This is because the delay of the compressor, composed of half adders (without carry chains), is independent of bit-width (shown in Table \ref{csa}).

\subsection{Reduction under the Same Bit-weight (OPT2)}

According to Eq. (\ref{eq6}), the \textbf{``shift"} is correlated with the $BW$. When rearranging loop unrolling, it's important to keep the \textbf{``shift"} within the $BW$ loop. This means restructuring the $BW$ as an outer loop that extends beyond the PE array while adjusting the \textbf{``shift"} in an outer loop. This positional transformation reduces the number of shifters and decreases the bit-width of subsequent components (compressor tree and DFFs) in PE, leading to a smaller area.

\begin{figure} [htbp]
  \flushleft 
  \includegraphics[scale=0.35]{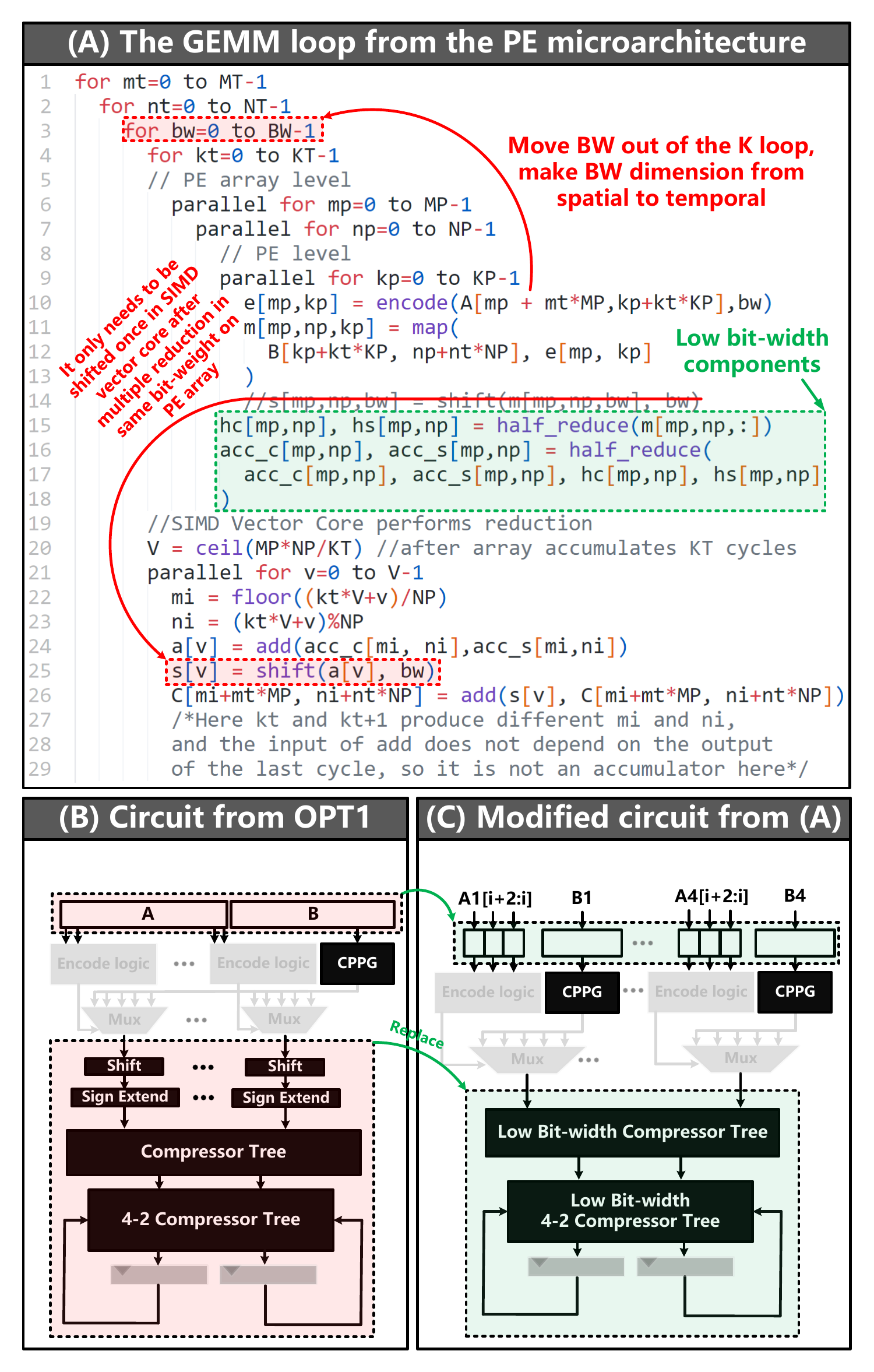}
  \caption{The proposed optimization architecture 2 (OPT2).}
  \label{opt2}
\vspace{-0.1cm}
\end{figure}

Please note that the $BW$ was initially adorned with ``parallel", indicating spatial unrolling in hardware. Simply reordering the $BW$ dimension to the outer level of the $K$ dimension will generate error reduction logic, as the \textbf{``half\_reduce"} is the reduction logic of $BW$ and needs to be at the same level as $BW$. When moving $BW$ to the outer level, its dimension should be transformed into a temporal dimension.

To maintain the throughput of the PE array, we partition the dimension $K$ into $K_P$ and $K_T$ (Figure \ref{opt2}(A) line 9 and 4), utilizing $K_P$ to fill the gaps in $BW$. Therefore, the \textbf{``half\_reduce"} in Figure \ref{opt2}(A) line 15 and 16 represents the reduction logic for dimensions $K_P$ and $K_T$, respectively. Similar to relocating the \textbf{``add"} in OPT1 (Sec. \ref{secopt1}), we can also transfer the \textbf{``shift"} to the SIMD vector core, requiring only a single shift after dimension $K_T$ has finished reduction.

After the \textbf{``shift"}, full adders are required to reduce the shifted PPs in order to ensure the correctness of the computation (Figure \ref{opt2}(A), line 26). The reason for using an \textbf{``add"} primitive here instead of \textbf{``accumulate"} is that the data stream from the PE array ensures that the indexes accessed by the SIMD vector core are unique from each other, and there are no accumulation dependencies within $K$ cycles in SIMD core. 
%Additionally, pipelined full adders can be implemented, but in Figure \ref{opt1}(A) line 14 should not use pipeline.

Therefore, pipelined full adders can be implemented here to boost frequency, while the pipeline design in Figure \ref{opt1}(A) is meaningless due to data dependencies.

With the shifter eliminated in the PE, the bit-width of input and output of \textbf{``half\_reduce"} in Figure \ref{opt2}(A) lines 15 $\sim$ 18 are also reduced, which further decreases the logic area (hardware architecture is shown in Figure \ref{opt2}(B)(C)).

However, there are two obvious drawbacks to this improvement. The first drawback is the increased bandwidth of PE. The second drawback is the potential increase in the number of CPPGs and input DFFs for operand B, which would occupy the additional area, for array designs, these additional areas can be shared among multiple PEs. And these drawbacks will be addressed step by step in the following subsections. In general, mapping the $BW$ to a temporal loop and reordering it to the outer loop of the $K$ dimension can reduce the area of shifters, compressor trees, and output DFFs. When considering the sparsity of encoding, temporal unrolling of the $BW$ could prove to be greatly advantageous.

\subsection{Acceleration with the Sparsity of Encoding (OPT3)}

In Sec. \ref{sec:motivation_sparsity}, we discussed the impact of zero bit-slices on operand encoding and their effect on average NumPPs in multiplication. Our proposed notation allows us to explore how encoding sparsity improves performance. Additionally, to address the drawback in OPT2, we introduce OPT3 in this section as a basis and fully resolve the issue with OPT4 in the next section. To describe the modified architecture, we introduce the \textbf{``sparse"} and \textbf{``sync"} primitives (in Table \ref{table6}).

The term \textbf{``sparse"} is used to compress inputs and obtain the indices of non-zero inputs. In contrast to previous work \cite{shi2024bitwave,im2024lutein,im2023sibia,lu2021distilling}, we use \textbf{``sparse"} for encoding numbers, while other works use it for multiplicands. 

\begin{table}[]\centering
\begin{tabular}{ll}
\midrule\midrule
\textbf{PRIMITIVE}       & \textbf{DESCRIPTION}                                                                                                                                                                     \\ \midrule[0.1pt]
$sparse(I_1, I_2,\dots,I_n)$ & \begin{tabular}[c]{@{}l@{}}  Outputs the indexes of non-zero inputs, \\ e.g. $[1,3] = sparse([0,1,0,2])$ .\end{tabular} \\ \midrule[0.2pt]
$sync()$                & \begin{tabular}[c]{@{}l@{}} Synchronizing sparse computation of \\ the PE subarrays.\end{tabular} \\ \midrule\midrule
\end{tabular}
\caption{Sparse and synchronization primitives.}
\vspace{0cm}
\label{table6}
\end{table}

\begin{figure} [htbp]
  \flushleft 
  \includegraphics[scale=0.31]{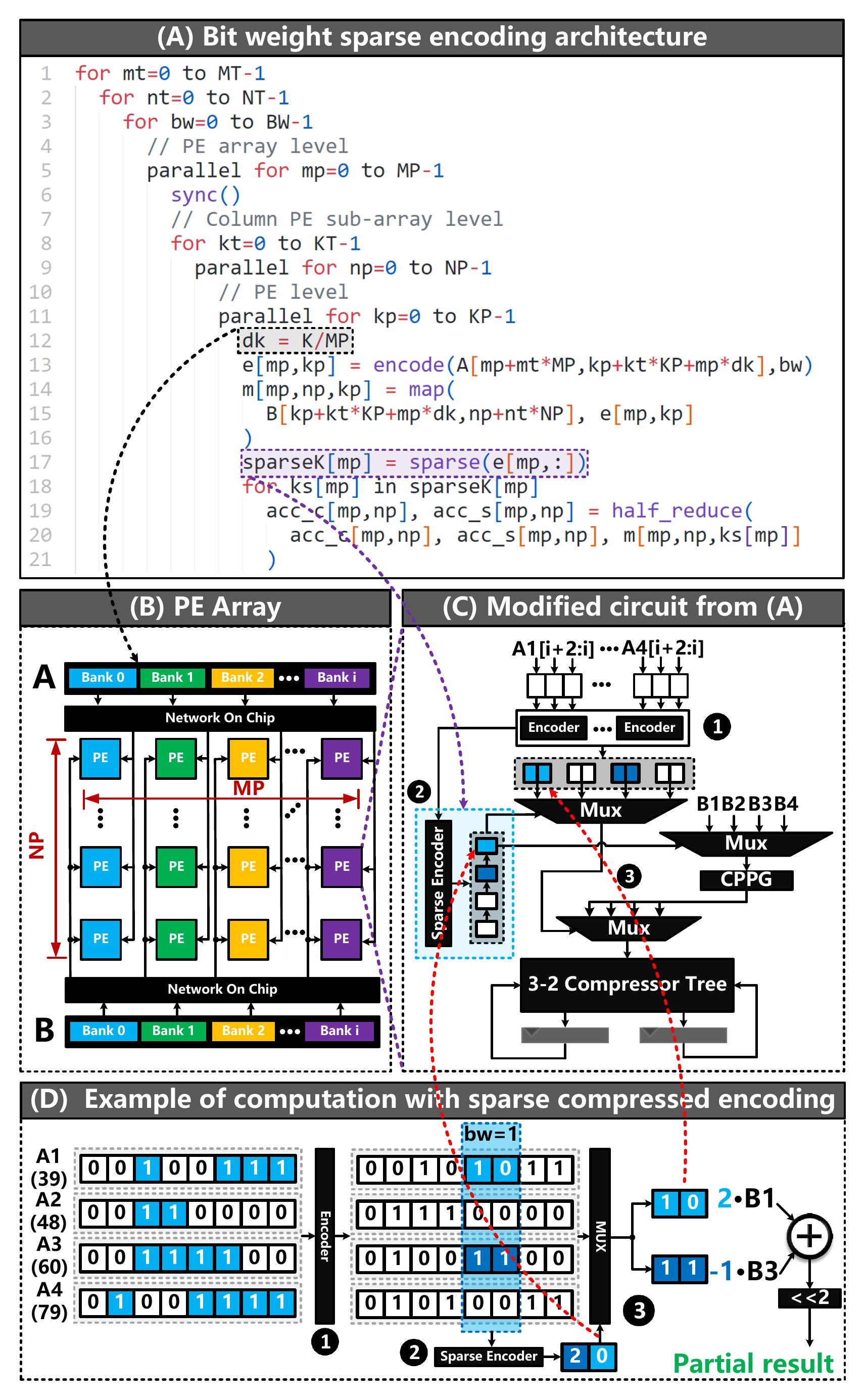}
  \caption{The proposed optimization architecture 3 (OPT3).}
  \label{opt3}
\vspace{0cm}
\end{figure}

To accumulate the non-zero PPs in reduction of $K$, we store the encoded number in the input DFFs of the PE (Figure \ref{opt3}(C) step \ding{182}). Then, an additional sparse encoder is introduced to output the non-zero index of the encoding number, as shown in Figure \ref{opt3}(C)(D) step \ding{183}. This index is then used as a selection signal for the non-zero PPs and multipliers B in step \ding{184}. Finally, a compressor is used to complete the accumulation. According to Table \ref{table2}, it takes only an average of 2.2 clock cycles to complete this equivalent multiplication, making the TPE more lightweight and able to run at higher frequencies.

\textbf{Since PEs in the same column can share the same multiplicand A in Figure \ref{opt3}(B), their computation time is uniform within a column but may differ across columns.} Therefore we introduce the \textbf{``sync"} to synchronize across PE columns. The \textbf{``sync"} blocks PE columns that finish earlier until all columns in the array are completed, indicating that the PE array is synchronized once every $T_{sync}$ cycle at most. $T_{sync}$ is determined by multiplying temporal loop cycles ($K_T$) and the number of unrolling operands in each PE ($K_P$), as unrolling operands are serialized to eliminate redundant multiplications by zero. For example, in Figure \ref{opt3}(A), column PEs are synchronized once at most every $K_T\times K_P$ cycles.

In Figure \ref{opt3}(A), we place the \textbf{``sync"} at the same level as the $K_T$ and below the spatial dimension $M_P$. This means that PEs in the same column share operand A. After $K_T$ iterations, the PE array synchronizes once. However, different columns work asynchronously, which can lead to bank conflicts. To avoid this, we switch the layout of A from $MK$ to $K_1 M_T K_2 M_P$, where $K_1$ and $K_2$ are sub-dimensions of the $K$ with sizes of $M_P$ and $K/M_P$. Similarly, the layout of B ($KN$) is mapped to $K_1 N_T K_2 N_P$. The elements of A with the same index in $K_1$ are stored in the same bank, and the index difference between two adjacent banks will be $dk$ in the $K$ dimension. So is the operand B. With this layout strategy, each column can access $N_P$ elements in A and $N_P$ elements in B in the same bank without data conflicts (Figure \ref{opt3}(A) line 12 $\sim$ 16). Here, we omit the primitive representation of the SIMD core.

Although computation time may vary for different PE columns, the total time will converge as long as there are sufficient elements along the $K$ dimension. To analyze the expected time between synchronizations, we define the number of non-zero PPs as a random variable $X$ follows $X \sim B(K,1-s)$ where the sparsity of encoding is $s$. Therefore, we can obtain the $\mu=K(1-s)$ and $\sigma=\sqrt{Ks(1-s)}$.

For the $M_P$ columns, let $T_i$ be the computation time for the $i$-th column when executing $K_T$ inputs. The interval between two \textbf{``sync"} operations is $T_{sync} = max(T_1, T_2, ..., T_{M_P})$. The $T_i$ is identically and independently distributed, and the cumulative distribution function $F(t)=P(T_{sync} \leq t)$ can be obtained as follows:

\begin{equation}
    \begin{aligned}
F(t)=\prod_{i=0}^{M_P} P(T_i \leq t)=\prod_{i=1}^{M_P}\sum_{j=0}^t \binom{K}{t}s^{K-j}(1-s)^{j}.
    \label{eq:c1}
    \end{aligned}
\end{equation}
Therefore, the mathematic expectation of $T_{sync}$ is:
\begin{equation}
    \begin{aligned}
E[T_{sync}]=\sum_{t=0}^{K}tP(T_{sync}=t)=K-\sum_{t=1}^{K-1}F(t).
    \label{eq:c2}
    \end{aligned}
\end{equation}

\begin{figure} [htbp]
  \flushleft 
  \includegraphics[scale=0.29]{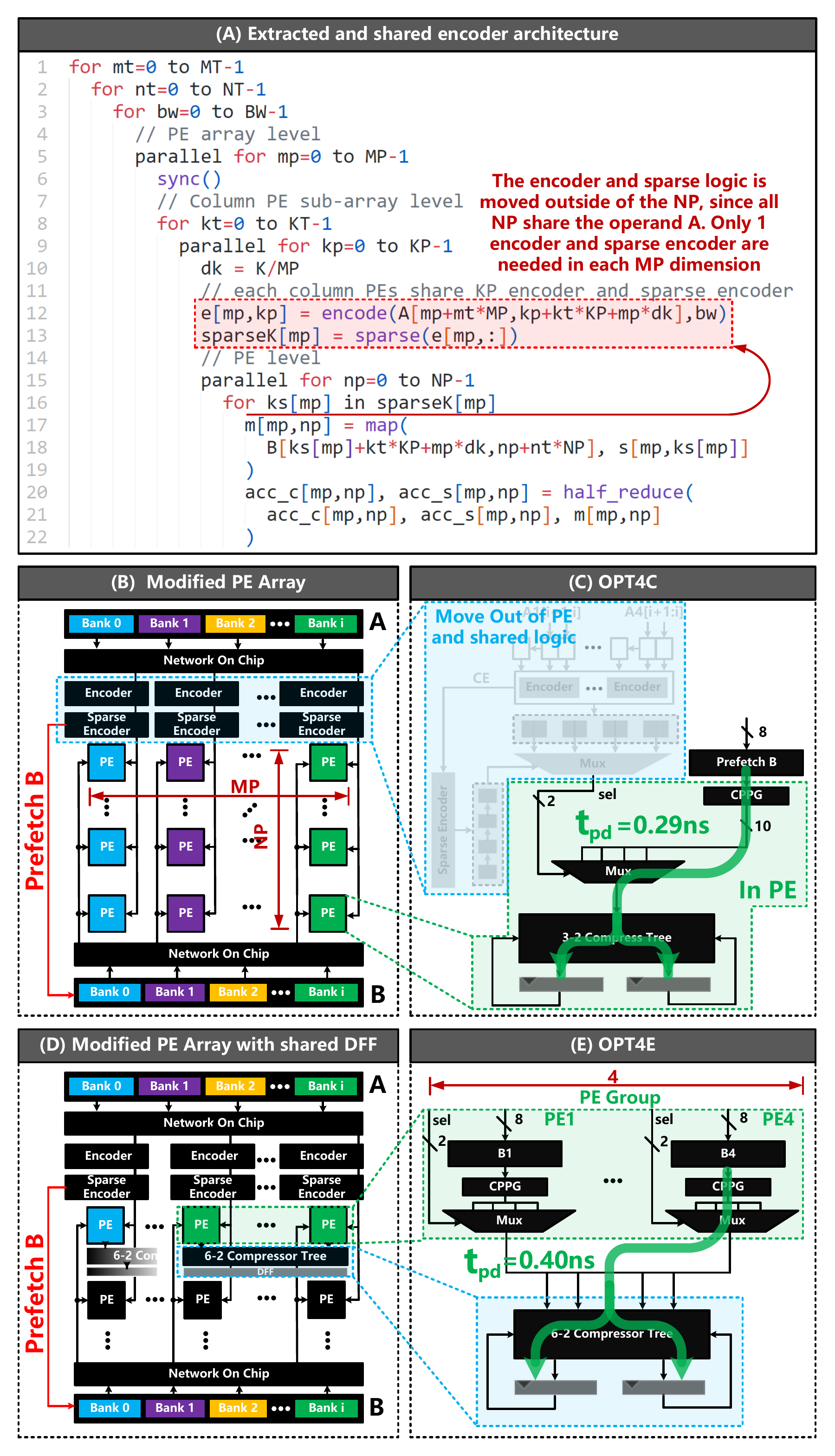}
  \caption{The proposed optimization architecture 4 (OPT4).}
  \label{opt4}
\vspace{-0.1cm}
\end{figure}

Based on Eq. (\ref{eq:c2}), when synchronization is performed at the granularity of the PE columns, acceleration based on encoding sparsity can result in an average reduction of $\sum_{t=1}^{K-1}F(t)$ cycles. In practical applications, such as DNN, the weights or inputs typically follow a normal distribution around zero. Taking a middle layer of ResNet-18 as an example and converting it into an MM through img2col, the reduction dimension size of the weights is 576 ($192 \times 3\times 3$). The sparsity of the weights is 0.38 when using EN-T\cite{wu2024entensorcore} for encoding. According to Eq. (\ref{eq:c1}) and Eq. (\ref{eq:c2}), the expected $T_{sync}$ would be 381, which represents a time saving of approximately 33.84\%.

In summary, directly analyzing the encoding number is more effective than multiplicand, as it is used to directly generate PPs, and skip consecutive ``1" bit-slices not only zero (e.g.``01111100$\rightarrow$10001100" in Figure \ref{back_1}). In OPT2, the PE computes PPs in parallel, requiring a compressor tree in the $K_P$. In contrast, OPT3 performs serial computations on the sparse compressed $K_P$ dimension. This change eliminates the need for a $K_P$-input compressor tree and transforms the 4-input compressor tree into a 3-input one in Figure \ref{opt3}(C). However, while this reduces the logic area, it does not address the increased bandwidth of operand B.  Additional optimizations are proposed in OPT4 below to efficiently address this issue.

\subsection{Extracted and Shared Encoder (OPT4C and OPT4E)}

Based on the analysis in Sec. \ref{secb}, we can rearrange the order of the $N_P$ and the $K_P$ (Figure \ref{opt4}(A) line 9 $\sim$ line 15), and move the \textbf{``encode"} and \textbf{``sparse"} to the outer level of $N_P$ dimension (here, we omit the primitive representation of the SIMD core.). Only the \textbf{``map"} and the \textbf{``half\_reduce"} remain in the innermost loop. Essentially, as operand A is broadcast across the PE columns, PEs in each column can share the same encoder and sparse encoder (Figure \ref{opt4}(B)). By placing the shared encoder outside the PE array, the duplication area of the encoder is reduced in each PE, which also reduces the bandwidth requirement of operand A. Additionally, with the sparse encoder located outside the PE array, the memory can recognize the sparsity of encoded operand A, and prefetch operand B by non-zero indices. With the out-of-plane encoder, the increased input in OPT2 is split and fed to PEs in a sequential manner. Each PE has access to only one shared encoding A and its corresponding prefetched B.

At this stage, PE contains only a CPPG, a Mux and a 3-2 compressor tree (Figure \ref{opt4}(C)), with a delay of only 0.29ns. The input ports of each PE include a 2-bit selection signal ``sel" and an 8-bit operand B, which reduces the bandwidth requirement. Moreover, compared to OPT3, it eliminates the encoding power consumption within each PE.

\begin{figure*} [htbp]
  \flushleft 
  \includegraphics[scale=0.14]{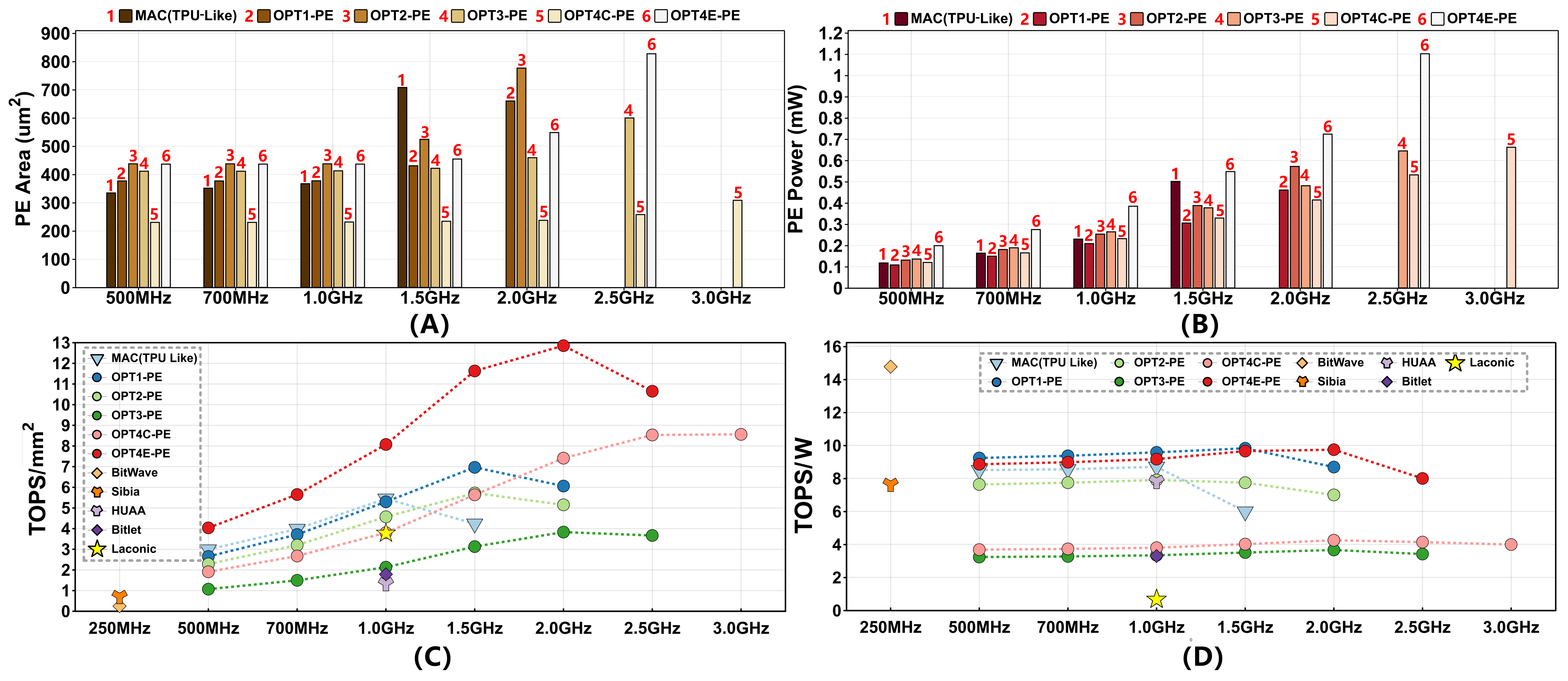}
  \caption{(A) PE area. (B) PE power consumption. (C) PE area efficiency curve. Under different clock constraints compared with the state-of-the-art. (D) PE energy efficiency curve. Under different clock constraints compared with the state-of-the-art.}
  \label{pe}
\vspace{0cm}
\end{figure*}

We propose an improved version with a higher computing density as shown in Figure \ref{opt4}(D)(E). We arrange 4 PEs in the same row into a PE group ($PE_g$), and the $PE_g$ shares one compressor tree and same DFFs. Four 3-2 compressor trees in $PE_g$ are merged into one shared 6-2 compressor tree. At this point, the external Encoder and Sparse Encoder of the PE Array, together with the CPPG in $PE_g$, form a non-zero partial product generator. This enables significant multiplication operation efficiency in large-scale MM, and the shared encoder also simplifies the internal logic of each PE, resulting in extremely low latency (easily up to 2GHz). Although there is a slight increase in the logical delay from 0.29ns to 0.40ns compared to Figure \ref{opt4}(C), this reduces the DFFs area and corresponding flip-flop power consumption in the PE Array by three-quarters, thereby improving the overall computational density and energy efficiency.

\section{Experiment}

\subsection{Experimental Setup}
\subsubsection{Hardware modeling}

\begin{figure} [htbp]
  \includegraphics[scale=0.15,center]{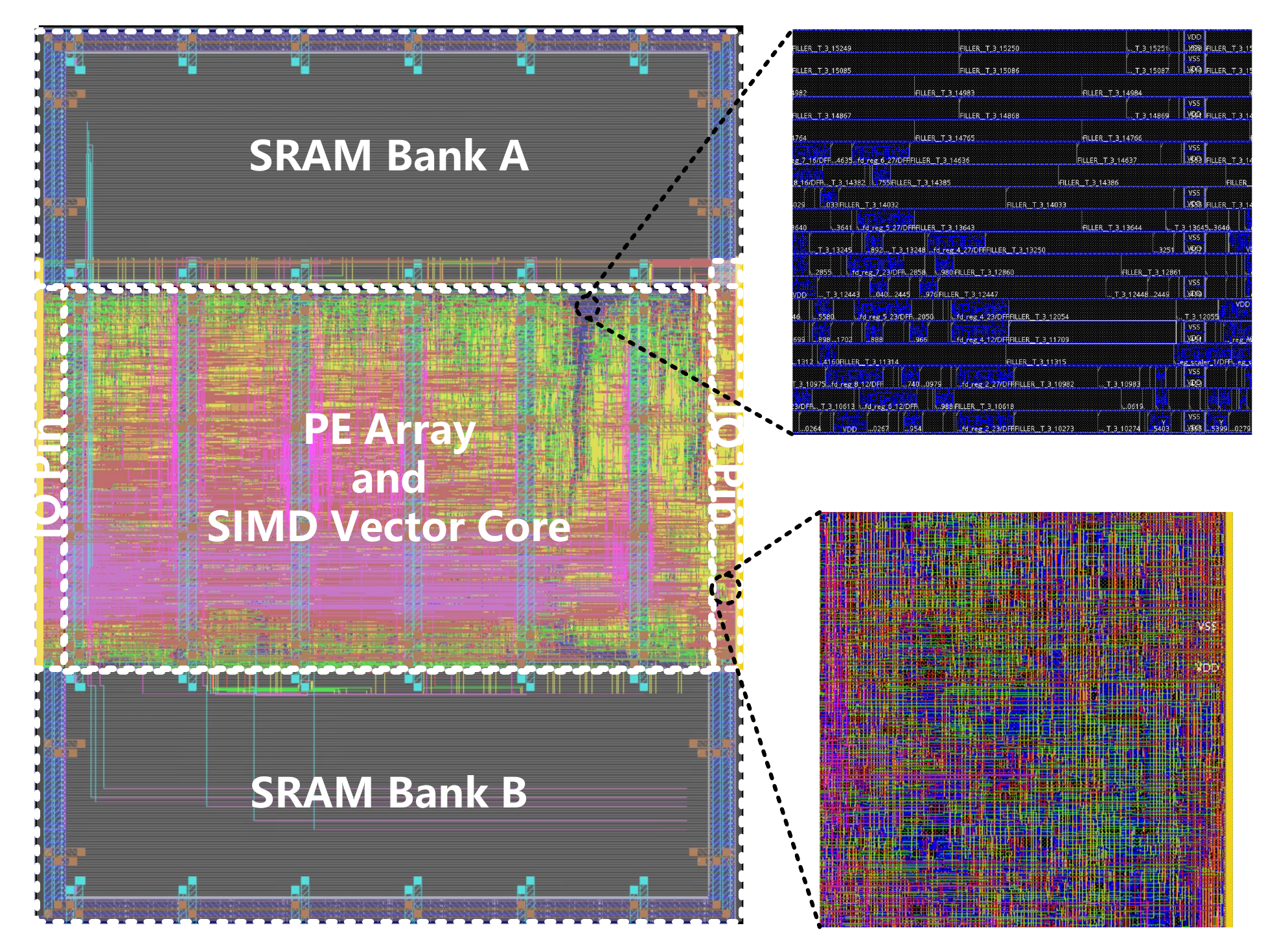}
  \caption{OPT4E chip layout (include IO and fillers).}
  \label{pr}
\vspace{-0.2cm}
\end{figure}

We implement our design in RTL and then synthesize it using the Synopsys Design Compiler with the SMIC 28nm-HKCP-RVT technology at an operating voltage of 0.72V. We utilize Cadence Innovus for placing and routing. Next, we use VCS to generate an FSDB waveform based on the given stimulus signals and use the waveform along with the optimized netlist, GEF and GDS files, the corresponding process corner, and physical libraries to evaluate hardware power consumption and timing using the PrimeTime PX tool. Finally, we employ Calibre to perform layout DRC/LVS checks. OPT4E chip layout is shown in Figure \ref{pr}.

\subsubsection{Experimental Arrangement}
In the second subsection, we evaluate the frequency characteristics of a single PE under given timing constraints, with a specific timing margin (8\% $\sim$ 10\% relative to the clock period). This evaluation includes five microarchitectures (OPT1, OPT2, OPT3, OPT4C, OPT4E) under INT8 MUL and INT32 ACC, which are compared with other PE microarchitectures (MAC (TPU-Like \cite{jouppi2023tpu}), Laconic \cite{sharify2019laconic}, Bitlet \cite{lu2021distilling}, Sibia \cite{im2023sibia}, Bitwave \cite{shi2024bitwave}, HUAA \cite{du202328nm}) under INT8. The benchmarks include area, power, area efficiency, and energy efficiency.
The test data consists of a normally distributed dense vector. The performance metric is the number of element-wise multiply-accumulate operations per second, and power is measured as the average power during the test. The area and power measurements include PE input/output DFFs, combinational logic, and clock networks.

In the third subsection, we test the dense matrix multiplication performance of the PE Array, using the same data distribution and performance-power testing methods as for the PE. Since the OPT1 and OPT2 are optimized for traditional PE arrays, the evaluations of the OPT1 and OPT2 are performed in four classic microarchitectures (systolic array (TPU \cite{jouppi2023tpu}), 3D-Cube (Ascend \cite{ascend}), multiplier-adder tree (Trapezoid \cite{Trapezoid}), and 2D-Matrix (FlexFlow \cite{lu2017flexflow})). The Cube contains 1000 ($10\times 10\times 10$) PEs, and others are $32\times 32$ PEs.
We also evaluate the performance of OPT3, OPT4C, and OPT4E ($32\times32 {PE_g}s$) in comparison with other bit-slice architectures.

\begin{table*}[]\centering
\tabcolsep=0.14cm 

\begin{threeparttable} 
\begin{tabular}{ccccccccc}
\midrule\midrule
\textbf{Others}                                                                                 & \textbf{TPU\tnote{$\diamondsuit$}}                                                  & \textbf{Ascend\tnote{$\diamondsuit$}}                                                  & \textbf{Trapezoid\tnote{$\diamondsuit$}}                                                  & \textbf{FlexFlow\tnote{$\diamondsuit$}}                                                  & \textbf{Laconic\tnote{$\blacklozenge$}}                                                   & \textbf{Bitlet\tnote{$\diamondsuit$}} & \textbf{Sibia\tnote{$\blacklozenge$}} & \textbf{Bitwave\tnote{$\diamondsuit$}} \\ \midrule
Frequency(MHz)                                                                         & 1000                                                          & 1000                                                             & 1000                                                                & 1000                                                               & 1000                                                               & 1000            & 250            & 250              \\
Area(um$^2$)                                                            & 370631                                                        & 320783                                                           & 283704                                                              & 332848                                                             & 213248                                                             & 415800          & 1069000        & 861681           \\
Power(W)                                                                                  & 0.25                                                          & 0.24                                                             & 0.22                                                                & 0.28                                                               & 1.21                                                               & 0.23            & 0.10           & 0.01             \\
Peak Performance(TOPS)                                                                 & 2.05                                                          & 2.05                                                             & 2.05                                                                & 2.05                                                               & 0.81                                                               & 0.74            & 0.77           & 0.22             \\
\begin{tabular}[c]{@{}c@{}}Energy Efficiency(TOPS/W)\end{tabular}                   & 8.05($\times$1.00)                                                  & 8.21($\times$1.00)                                                     & 9.31$\times$1.00)                                                        & 7.29($\times$1.00)                                                       & 0.67($\times$1.00)                                                        & 3.29($\times$4.91)     & 7.65($\times$11.42)   & 14.77($\times$22.04)    \\
\begin{tabular}[c]{@{}c@{}}Area Efficiency(TOPS/mm$^2$)\end{tabular} & 5.53($\times$1.00)                                                   & 7.22($\times$1.00)                                                      & 7.22($\times$1.00)                                                         & 6.15($\times$1.00)                                                        & 3.77($\times$1.00)                                                        & 1.79($\times$0.47)     & 0.72($\times$0.19)    & 0.25($\times$0.07)      \\ \midrule
\textbf{Ours}                                                                                   & \textbf{\begin{tabular}[c]{@{}c@{}}OPT1\tnote{$\diamondsuit$}\\ (TPU)\end{tabular}} & \textbf{\begin{tabular}[c]{@{}c@{}}OPT1\tnote{$\diamondsuit$}\\ (Ascend)\end{tabular}} & \textbf{\begin{tabular}[c]{@{}c@{}}OPT1\tnote{$\diamondsuit$}\\ (Trapezoid)\end{tabular}} & \textbf{\begin{tabular}[c]{@{}c@{}}OPT1\tnote{$\diamondsuit$}\\ (FlexFlow)\end{tabular}} & \textbf{\begin{tabular}[c]{@{}c@{}}OPT2\tnote{$\diamondsuit$}\\ (FlexFlow)\end{tabular}} & \textbf{OPT3\tnote{$\diamondsuit$}}   & \textbf{OPT4C\tnote{$\diamondsuit$}} & \textbf{OPT4E\tnote{$\blacklozenge$}}   \\ \midrule
Frequency(MHz)                                                                         & 1500                                                          & 1500                                                             & 1500                                                                & 1500                                                               & 1500                                                               & 2000            & 2500           & 2000             \\
Area(um$^2$)                                                            & 436646                                                        & 332185                                                           & 271989                                                              & 373898                                                             & 347216                                                             & 460349          & 259298         & 672419           \\
Power(W)                                                                                  & 0.37                                                          & 0.24                                                             & 0.22                                                                & 0.38                                                               & 0.35                                                               & 0.70            & 0.51           & 0.89             \\
Peak Performance(TOPS)                                                                 & 3.07                                                          & 3.07                                                             & 3.07                                                                & 3.07                                                               & 3.07                                                              & 1.80            & 2.25           & 7.22             \\
\begin{tabular}[c]{@{}c@{}}Energy Efficiency(TOPS/W)\end{tabular}                   & 8.41($\times$1.04)                                                   & 12.82($\times$1.56)                                                     & 13.89($\times$1.49)                                                        & 8.08($\times$1.11)                                                        & 8.77($\times$1.20)                                                        & 2.57($\times$3.83)     & 4.41($\times$6.58)    & 8.11($\times$12.10)     \\
\begin{tabular}[c]{@{}c@{}}Area Efficiency(TOPS/mm$^2$)\end{tabular} & 7.04($\times$1.27)                                                   & 9.25($\times$1.28)                                                      & 11.29($\times$1.56)                                                        & 8.22($\times$1.34)                                                        & 8.85($\times$1.44)                                                        & 3.91($\times$1.04)     & 8.68($\times$2.30)    & 10.73($\times$2.85)  \\ \midrule\midrule
\end{tabular}
         \begin{tablenotes} 
        \footnotesize            
        \item[$\diamondsuit$]Reports on timing, power, and area after logic synthesis.        
        \item[$\blacklozenge$]Reports on timing, power, and area after placing and routing by chip layout.            
      \end{tablenotes}            
 \end{threeparttable}
\caption{Comparision with state-of-the-art in PE Array level on matrix multiplication.}
\vspace{0cm}
\label{array}
\end{table*}

\subsection{PE Comparision}
\subsubsection{Area and area efficiency}

\subsubsection{Comparison Method}
For computation arrays in TPU, Ascend, Trapezoid, and FlexFlow, we utilize Verilog HDL to recurrent their design. In the case of bit-slice architectures such as Laconic, Bitlet, Sibia, and Bitwave, we extract the area and power breakdowns of the PE arrays from their respective papers. When dealing with process nodes other than 28nm, the results are normalized to the 28nm process for performance comparison. The conversion methods for process and power are based on references from TSMC ANNUAL REPORT \cite{tsmc2023}.

At 28nm, reaching 1GHz represents a performance inflection point for traditional MAC (TPU-Like). However, due to the constraints of the high bit-width accumulator, it is nearly impossible to maintain a comparable area while under the 0.63 ns clock constraint. As a result, when running at 1.5GHz, traditional MAC experiences a significant increase in area (as illustrated in Figure \ref{pe}(A)), growing from 367$um^2$ to 707$um^2$. At this point, the synthesis tool replicates a large amount of logic within the MAC to maintain parallelism and reduce latency. Consequently, for traditional MACs, surpassing 1GHz does not lead to further improvements in area efficiency (as depicted in Figure \ref{pe}(C)).

In contrast, the latency is independent of bit width in half-adder accumulation schemes (OPT1 $\sim$ OPT4). Therefore, our proposed designs can operate at frequencies above 1.5GHz and achieve high area efficiency. When constrained from 1.0 GHz to 1.5 GHz, the synthetic area of OPT1 is only increased by a factor of 1.14, compared to a factor of 1.93 in TPU-like MACs. This represents a significant improvement in the area efficiency of OPT1 in 1.5GHz.

OPT2 exhibits a similar timing trend but with an increase in area. While OPT2 reduces the area of the reduction logic and output DFFs, it does so at the expense of increased PE bandwidth and it is essential to consider the additional increase in area and power consumption of input DFFs. As a result, OPT2 doesn't offer an advantage over a single PE. However, there are various ways to reduce the average input width of the DFFs in the array level, such as local broadcast and local shared DFFs. Therefore, OPT2 can achieve optimization by sharing input DFFs among multiple PEs through local broadcasting.

OPT3 skips zero PPs. In terms of area analysis for a single PE, similar to OPT2, the inclusion of input DFFs for multiple operands results in a significant occupation of area in a single PE. However, the area and delay of combinatorial logic are significantly reduced. When constrained from 1.5 GHz to 2.0 GHz, the synthetic area of OPT3 increases only by a factor of 1.09 with a peak frequency of 2.5 GHz. From the analysis of area and frequency, it can be concluded that the inflection point of area efficiency performance for OPT3 is above 2.0GHz, and the use of the pre-fetch mechanism in OPT4 effectively addresses the area issue of input DFFs through an external encoder. 

In comparison to other bit-slice architectures, OPT3 maintains an area efficiency (in 2GHz) on par with Laconic, 2.12 times that of Bitlet, 5.28 times that of Sibia, and 15.2 times that of Bitwave. Most of these architectures operate at clock frequencies ranging from 250MHz to 1GHz.

% By studying these microarchitectures, it can be observed that bit-serial algorithms typically perform 1-bit or 2-bit parallel multiplications, resulting in extremely low logic area and latency. 
It can be observed that bit-serial algorithms typically perform 1-bit or 2-bit parallel multiplications, resulting in extremely low logic area and latency. 
However, the reduction and accumulation of PPs cause bottlenecks in these architectures, leading to peak frequencies similar to MAC (1GHz). Thus, the key to improving the computational density is to replace the reduction logic with a lighter-weight alternative.

OPT4C and OPT4E are optimizations of OPT3 at the array level. Through sharing 2 encoders among PE columns, making PEs more lightweight, and reducing the input DFFs area through sparse coding prefetching operations. These improvements further enhance the area efficiency compared to OPT3. In addition, OPT4E aims to balance the area ratio between DFFs and joint logic to achieve high area efficiency.

\begin{figure*} [htbp]
  \flushleft 
  \includegraphics[scale=0.25]{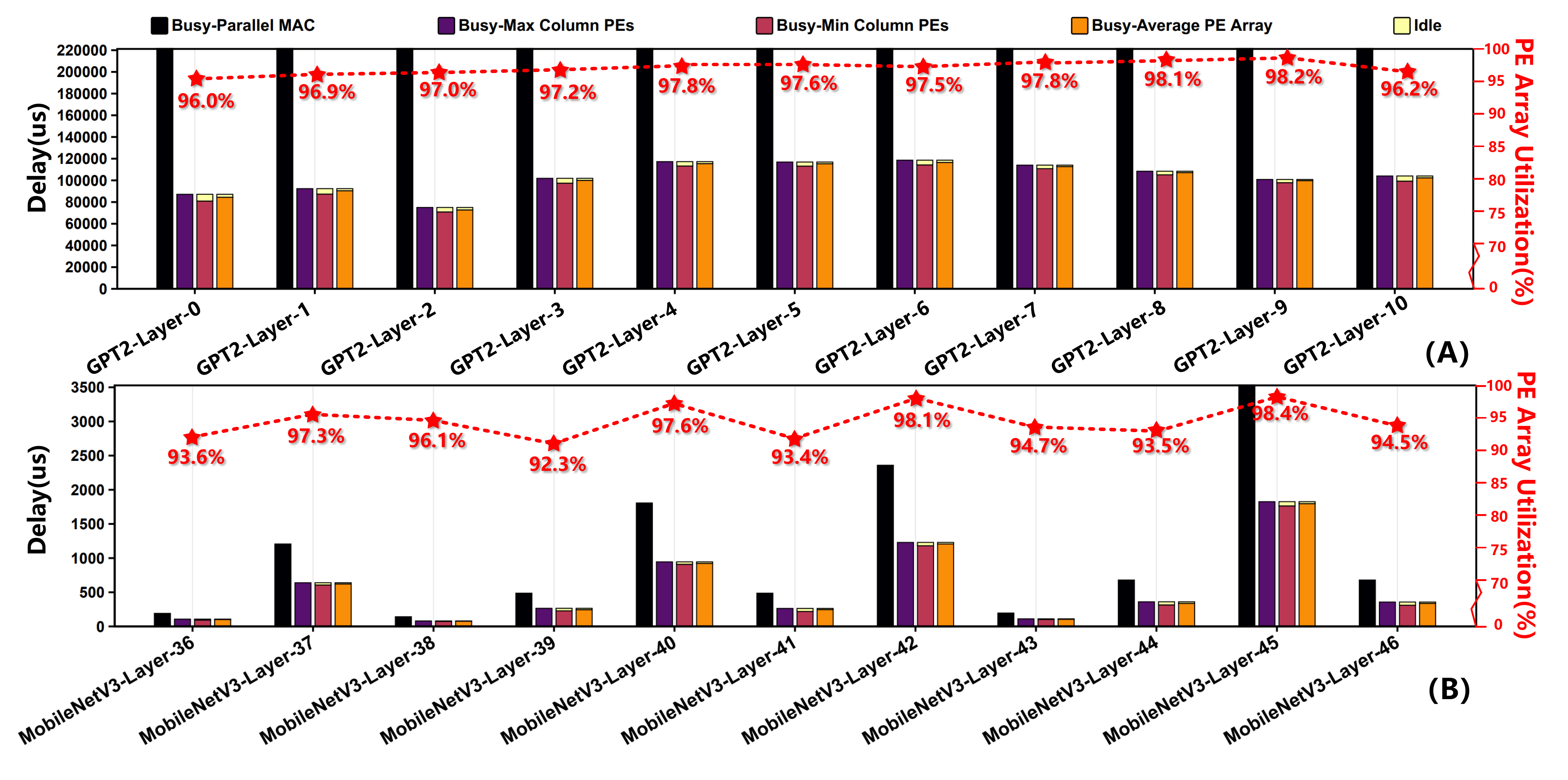}
  \caption{Comparison of the computational performance of a single tile in the TPEs composed of OPT4E and parallel MAC under (A) GPT-2 layer and (B) MobileNetV3 sub-layers, along with an analysis of the utilization of OPT4E array.}
  \label{11}
\vspace{0.3cm}
\end{figure*}

\begin{figure} [htbp]
  \flushleft 
  \includegraphics[scale=0.16]{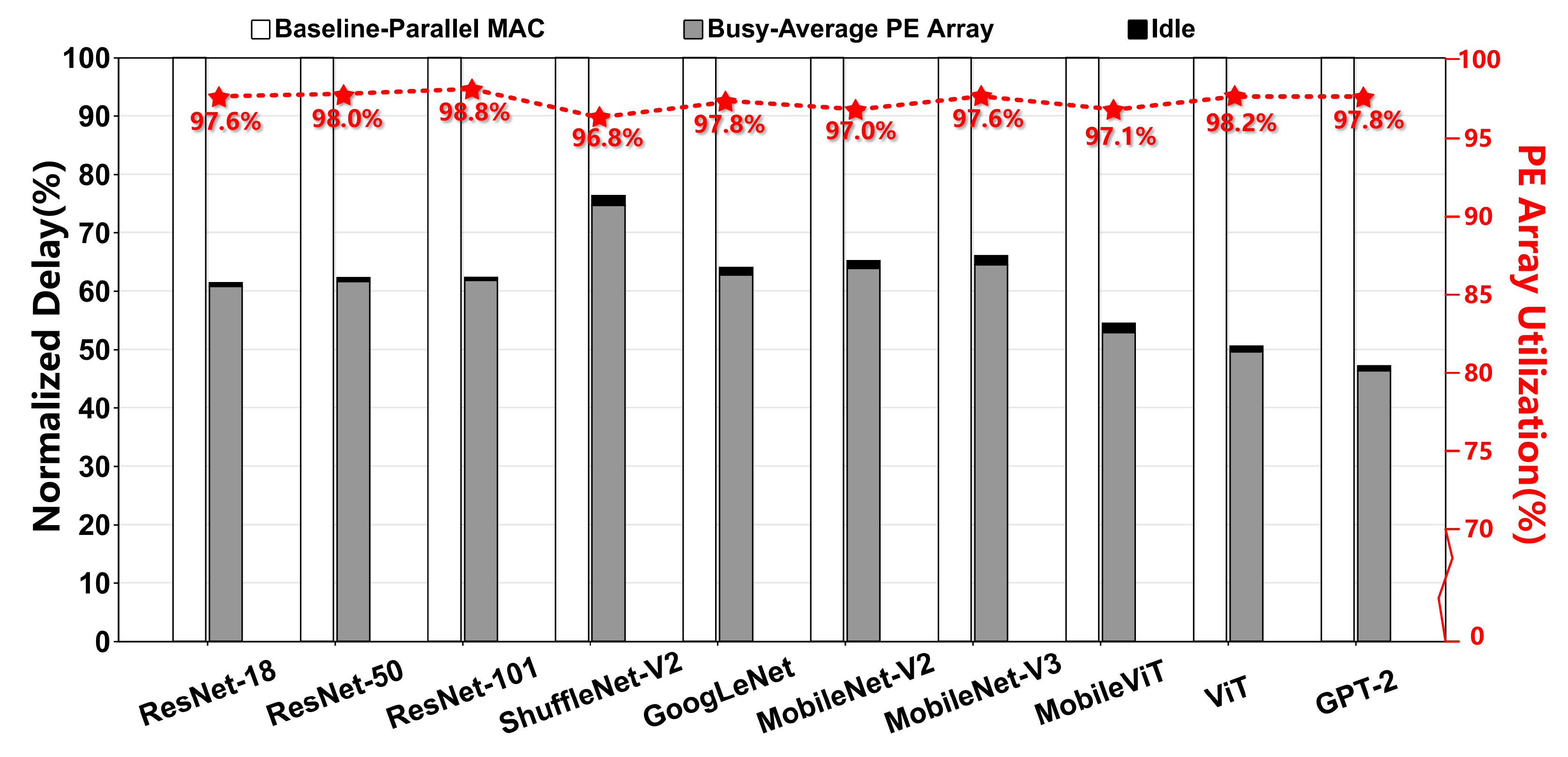}
  \caption{Comparison of performance of TPEs normalization composed of OPT4E and parallel MACs under different networks, and the total idle ratio of OPT4E subarrays.}
  \label{12}
\vspace{-0.1cm}
\end{figure}

\subsubsection{Power and energy efficiency}

%  \begin{figure*} [htbp]
%   \flushleft 
%   \includegraphics[scale=0.219]{array.pdf}
% \caption{Comparison of optimized architecture with corresponding benchmarks and other configurations. (A) Area efficiency up ratio. (B) Energy efficiency up ratio on traditional architecture. (C) Energy efficiency up ratio on bit-slice architecture. (D) INT16 MUL and INT8 ACC with OPT1 scheme. (E)(F) Scalability testing of the OPT4E array.}
%   \label{up}
% \vspace{-0.1cm}
% \end{figure*}

The significant impact of the DFFs and clock network on power analysis can't be overlooked. In high-speed digital circuits, the clock network accounts for 30\%$\sim$60\% of total power consumption, leaving 40\%$\sim$70\% to be optimized by logic designers, including DFFs and combinational logic power consumption. Despite optimizations in logic regions such as OPT1, OPT2, and OPT3, the increase in clock network power consumption at high frequencies exceeds the increase in combined logic power consumption. Therefore, when the frequency increases to a certain threshold, the energy efficiency will decrease.

Designers can reduce power consumption at high frequencies by minimizing the register area within the logic design. This consideration is reflected in designs such as OPT4C and OPT4E, which reduces the need for input and output DFFs while balancing the logic and DFFs regions. Ultimately, OPT4E enables significant computational density while maintaining energy efficiency, as demonstrated in Figure \ref{pe}(B) and (D).

\subsection{Array-level comparison with state of the art}

\subsubsection{Configuration setup}

In the experimental deployment of the PE array, since EDA tools require constraint files to be read before synthesis, it is necessary to use predefined delays to constrain the clock. To this end, we thoroughly tested the frequency range for each PE design, as shown in Figure 9, aiming to determine the optimal clock frequency for each configuration (achieving better energy and area efficiency).

From a detailed observation of Figure \ref{pe}(A), the frequency range of the TPU-like MAC spans from 500 MHz to 1.5 GHz. Beyond 1.5 GHz, timing violations occur, preventing normal operation. Only design 5 (OPT4C) can reach 3.0 GHz, but higher frequencies do not always lead to better synthesis performance.

As shown in Figures \ref{pe}(C) and \ref{pe}(D), the TPU-like MAC-based design achieves peak area and energy efficiency at 1.0 GHz. The frequency limit of the PE using the OPT1 design is 2.0 GHz, but its synthesis performance is optimal at 1.5 GHz. Similarly, we identified the optimal frequency points for OPT3, OPT4C, and OPT4E, which were then used as clock constraints for synthesizing and testing the TPEs.

\begin{figure*} [htbp]
  \flushleft 
  \includegraphics[scale=0.17]{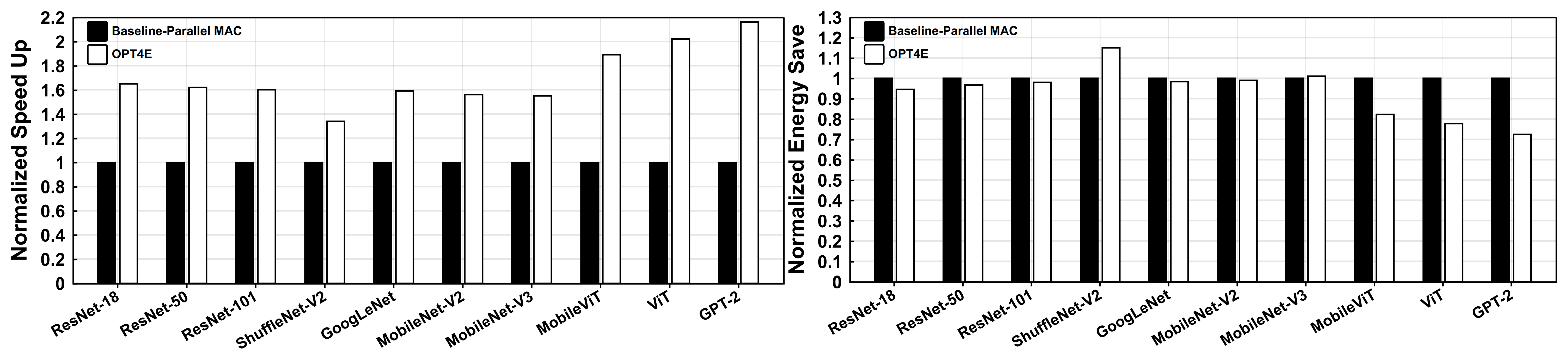}
  \caption{The normalized speedup and energy consumption ratio of TPEs composed of OPT4E and parallel MAC.}
  \label{13}
\vspace{-0.1cm}
\end{figure*}

\begin{figure} [htbp]
  \flushleft 
  \includegraphics[scale=0.18]{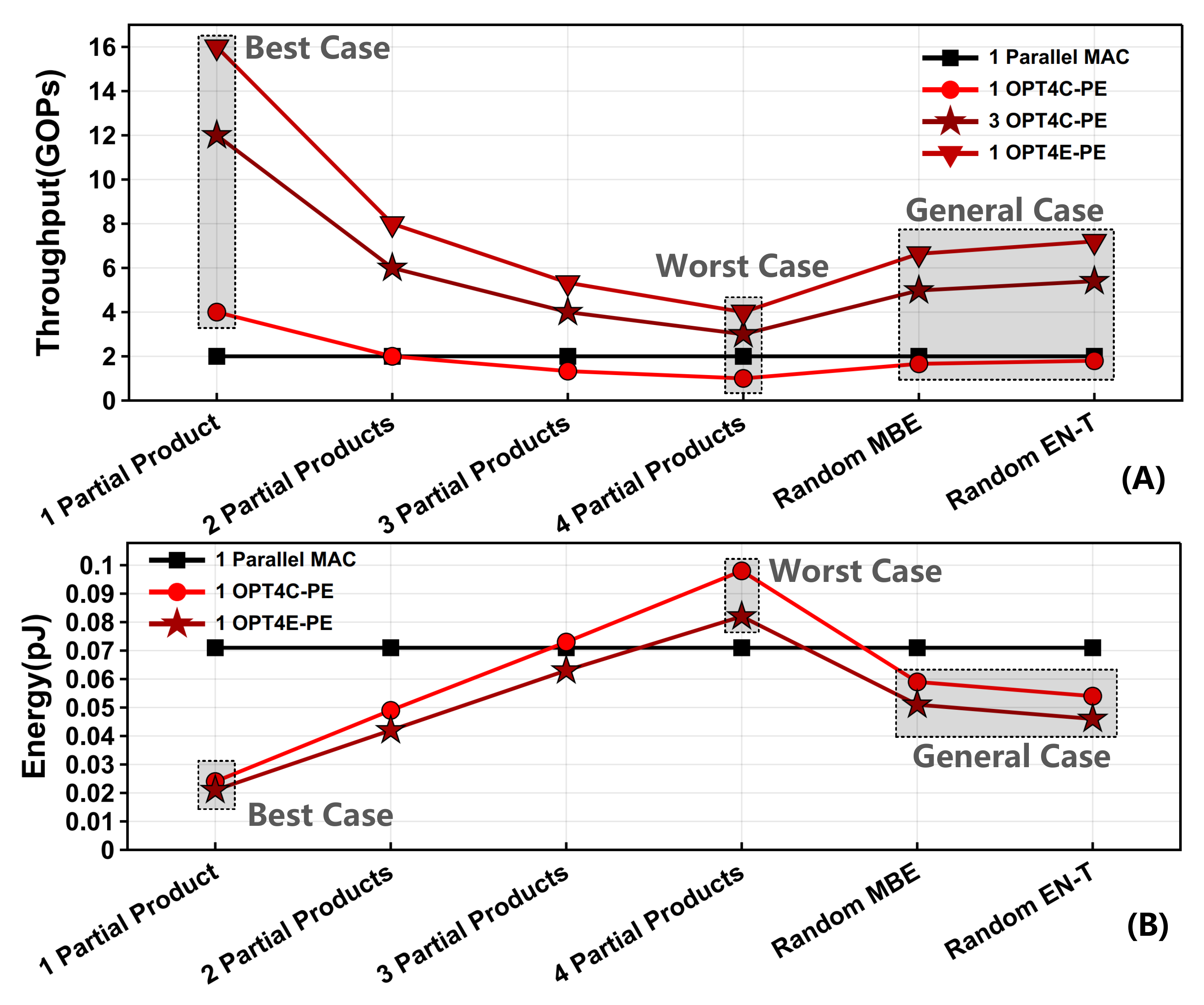}
  % \caption{(A) Throughput for different PEs. 1 Parallel MAC (246$um^2$) $\approx$ 3 OPT4C-PE (81.27$um^2$) $\approx$ 1 OPT4E-PE (311$um^2$). (B) Energy consumed per multiplication-accumulation operation.}
  \caption{(A) Throughput for different PEs. 1 Parallel MAC (246$um^2$) $\approx$ 3 OPT4C-PE (81.27$um^2$) $\approx$ 1 OPT4E-PE (311$um^2$). (B) Energy consumed per multiplication-accumulation operation.}
  \label{14}
\vspace{-0.2cm}
\end{figure}

\subsubsection{Comparision with classical TPE architecture}

As depicted in Table \ref{array}, we implement the OPT1 on conventional architectures such as TPU (systolic array), Ascend (3D-Cube), Trapezoid (multiplier-adder tree), and FlexFlow (2D-Matrix). For FlexFlow (2D-Matrix), OPT2 is employed. Subsequently, we compare the performance enhancements before and after applying these optimizations, using them as benchmarks. Based on our previous analysis of the area efficiency per PE, we observe an increase in area efficiency across all four microarchitectures, by a factor of 1.27, 1.28, 1.58, 1.34 and 1.44, respectively. Energy efficiency was increased by 1.04, 1.56, 1.49, 1.11 and 1.20 times, respectively. Moreover, for the OPT2 particularly in FlexFlow (2D-Matrix), there was a slight improvement over OPT1 which aligns with our previous analysis of PE area efficiency. The reason for this improvement is that the 2D-Matrix architecture broadcasts inputs across its rows and columns, allowing a single input DFFs to be shared among PE rows and columns, leading to a dilution of the area of OPT2's input registers, demonstrating the advantage of having lower bit-widths within PEs.

\subsubsection{Comparison with the bit-slice architecture}

Choosing Laconic as the comparison baseline for bit-slice architecture (Bitlet, Sibia, BitWave, OPT3, OPT4C, and OPT4E in Table \ref{array}) reveals a common trait: these methods typically improve energy efficiency significantly but generally lack in area efficiency. Despite their compact size, they aren't as computationally efficient, making it challenging to significantly increase the computational power per unit area. In terms of computational efficiency, the bit-slice technique can be improved in two main ways. First, the number of PPs is reduced by sparse encoding. Second, eliminate the bottleneck of accumulation in bit-slice operations. Hence, the optimization strategy of OPT1 led to the development of OPT3, which effectively addresses these issues and is also applicable to bit-serial processing. Additional advancements include higher-level loop optimizations in arrays such as operand sharing enabling us to propose encoding within all bit-slice PEs to further reduce area and improve timing. Additionally considering the balance between combinational logic and DFFs is a crucial step toward further reducing area efficiency and energy consumption. Finally, our final iteration OPT4E not only maintains commendable energy efficiency but also significantly enhances the computational density of bit-slice architecture.

\subsection{Workloads for DNNs and LLMs}
Unlike the TPEs formed by parallel MACs, the throughput of MM provided by the OPT4E is mainly influenced by two factors: (1) The number of partial products after encoding of the multiplicand; 
% (2) the vector dimension of the reduction. 
(2) the reduction dimension of the vector. 

As shown in Figure \ref{14}(A), the throughput of parallel MACs is not affected by the number of partial products of the operands. 
% Regardless of the number of partial products of the multiplicand, traditional MACs will parallelly reduce four partial products, resulting in constant computational power and energy consumption as shown in Figure \ref{14}(B). 
The traditional MACs always parallelly reduce 4 partial products, resulting in constant computational power and energy consumption as shown in Figure \ref{14}(B). 
In contrast, the area of a single PE in the OPT4C (81.27$um^2$) is about one-third of the parallel MAC (246$um^2$). In the best-case scenario, all inputs produce only one partial product after encoding, achieving twice the throughput of a regular MAC with one-third energy savings. In the worst-case scenario, all inputs produce 4 partial products after encoding, resulting in an equivalent computational power of half that of a regular MAC. 
In more general cases, for a set of normally distributed vectors, the average number of partial products for MBE and EN-T encoding is 2.41 and 2.22 respectively (as shown in Table \ref{table2}). Therefore, a single OPT4C can achieve a throughput close (1.8 GOPS) to that of a regular MAC with lower energy consumption. When comparing equal areas, we used three OPT4Cs and one OPT4E, which generally achieve 2.7$\times$ and 3.6$\times$ the throughput improvement compared to parallel MACs, with lower energy consumption per operation. Even in the worst case, a certain speedup can still be achieved.

The second factor influencing throughput is the dimensionality of the reduction vector. Since synchronization among different column PEs in OPT4E is necessary after the reduction is completed, a higher vector dimensionality leads to reduced variance in computation time across the column PEs, resulting in improved performance. To illustrate this with practical deep neural networks (DNNs), we select two representative NN layers: the Transformer layer of GPT-2 and the Depthwise (DW)-Pointwise (PW) layer of MobileNet, as depicted in Figures \ref{11} and \ref{12}. We employe a systolic array and the OPT4E architecture of the same area for comparing inference delays. We record the fastest computing column PEs (Busy-Min Column PEs), the slowest computing column PEs (Busy-Max Column PEs), and the average busy and idle ratios of all column PEs (Busy-Average PE) for comparison. The specific meaning of delay in Figure \ref{11} refers to the time required for vector reduction under a single excitation (e.g., in a GPT layer, delay represents the inference latency of a single embedding vector at each layer, while for MobileNet, delay refers to the inference time of a single pixel at each layer).

In the multi-head attention layer of GPT-2, which typically involves higher-dimensional matrix multiplication, the idle time has minimal impact on overall computational efficiency. In contrast, MobileNetV3 exhibits a lower reduction dimension in the DW layer and a higher reduction dimension in the PW layer, resulting in lower utilization in the DW layer compared to the PW layer. However, since the computational load of the PW layer is significantly greater than that of the DW layer, a notable speedup can still be achieved across all layers.

As illustrated in Figures \ref{12} and \ref{13}, we compared the inference performance of several mainstream backbones. MobileVIT, VIT, and GPT-2 achieved the highest speedup ratios, with performance improvements of 1.89, 2.02, and 2.16 times, respectively. Regarding energy consumption, as shown in Figure \ref{13}, networks with higher reduction dimensions tend to achieve greater energy savings.

% \subsection{Other Configurations}

% Further, we evaluate the performance of OPT1 in INT16 MUL and INT48 ACC, as shown in Figure \ref{up}(D). Traditional MAC typically face bottlenecks at higher bit-widths accumulation. However, OPT1 shows significant improvement by reducing the latency dependence on the accumulation bit-width. Instead, the max frequency depends primarily on the multiplier width, allowing OPT1 to efficiently maintain both frequency and area efficiency.  Additional exploration into the scalability of OPT4E (Figure \ref{up}(E) and (F)) spans across array sizes (32$^2$, 64$^2$, 128$^2$, and 256$^2$) for comprehensive synthesis and power evaluation. Our findings show that OPT4E maintains commendable energy and is efficiency up to a 64$^2$ array size. However, as the array size increases, input timing faces more pressure, impacting overall performance. The size of subarrays in existing accelerators is not very large (less than 128$^2$) to avoid affecting utilization \cite{norrie2020google}.

\section{Disscussion}

In actual calculations, column PEs may experience idle times due to early completion of computations. The occurrence of idle periods (bubbles) in column PEs benefits TPEs, as PEs handling vectors with fewer non-zero partial products can quickly complete computations and enter an idle state, saving power. However, processing performance depends on the slowest column PEs. For matrices with higher vector dimensions, the variance in the reduction clock cycles across column PEs gradually decreases. Consequently, as the computation load increases, the bubble ratio also declines. This results in significant benefits from both power consumption and computation speed perspectives. For OPT1 design, all PEs are synchronized throughout the entire computation process. Conversely, for sparse computations encoded in OPT3, OPT4C, and OPT4E, not every MAC clock cycle differs. Since the same multiplicand is broadcast to all column PEs, the reduction cycle for each column PE is identical.

Overall, the MAC plays a critical role in determining the area and performance of AI DSA. Analyzing the MAC components enables the identification of area and delay bottlenecks in each subcomponent, which indirectly impacts TPE performance. In Section \ref{secb}, we discuss the validity of component position transformations within nested loops, facilitating the exploration of higher-dimensional transformations in the search space. Furthermore, selecting encoded operands represents an additional dimension in the search space. Prioritizing operands with high sparsity enhances acceleration, further broadening the optimization search space.

\section{Conclusion}

Traditional TPE designs primarily focus on data flow reuse through MAC-based specialized matrix multiplication units. This work extends TPE design to the component level within the MAC, identifying bottlenecks through higher-level loop transformations. We first introduce a fine-grained primitive that uncovers a broader design space for TPE. Within this expanded space, we analyze bottlenecks by exposing the implicit dimensions of traditional MAC designs. Subsequently, we apply valid loop transformations across components to address these bottlenecks, resulting in more efficient parallel hardware and providing a methodology for designing high-performance PE microarchitectures. Furthermore, we investigate the principles of bit-sparsity acceleration, where encoded multiplicands enhance operand sparsity, allowing the elimination of zero partial products to achieve sparse acceleration. Leveraging the proposed primitives, we develop a TPE microarchitecture that compresses non-zero partial products, significantly improving performance.

\end{document}